\documentclass[sigconf]{acmart}
\usepackage{makecell}
\usepackage[capitalise, noabbrev]{cleveref}
\usepackage{booktabs}
\usepackage{booktabs}
\usepackage{ifthen}
\newboolean{highlightversion}
\setboolean{highlightversion}{false}  

\newcommand{\revise}[1]{%
    \ifthenelse{\boolean{highlightversion}}%
    {\textcolor{red}{#1}}  
    {#1}                   
}

\AtBeginDocument{%
  }

\acmConference[UIST '25]{The ACM Symposium on User Interface Software and Technology (UIST)}{Sep 28--Oct 01, 2025}{Busan, Korea}
\copyrightyear{2025}
\acmYear{2025}
\setcopyright{cc}
\setcctype{by}
\acmConference[UIST '25]{The 38th Annual ACM Symposium on User Interface Software and Technology}{September 28-October 1, 2025}{Busan, Republic of Korea}
\acmBooktitle{The 38th Annual ACM Symposium on User Interface Software and Technology (UIST '25), September 28-October 1, 2025, Busan, Republic of Korea}\acmDOI{10.1145/3746059.3747750}
\acmISBN{979-8-4007-2037-6/2025/09}

\acmSubmissionID{7127}

\begin{document}

\title[ChairPose]{ChairPose: Pressure-based Chair Morphology Grounded Sitting Pose Estimation through Simulation-Assisted Training}

\author{Lala Shakti Swarup Ray}
\affiliation{%
  \institution{DFKI, RPTU}
  \city{Kaiserslautern}
  \country{Germany}
  }
\email{lala_shakti_swarup.ray@dfki.de}
\orcid{0000-0002-7133-0205}

\author{Vitor Fortes Rey}
\affiliation{%
  \institution{DFKI, RPTU}
  \city{Kaiserslautern}
  \country{Germany}}
\email{mengxi.liu@dfki.de}

\author{Bo Zhou}
\affiliation{%
  \institution{DFKI, RPTU}
   \city{Kaiserslautern}
  \country{Germany}}
\email{bo.zhou@dfki.de}

\author{Paul Lukowicz}
\affiliation{%
  \institution{DFKI, RPTU}
  \city{Kaiserslautern}
  \country{Germany}}
\email{paul.lukowicz@dfki.de}

\author{Sungho Suh}
\authornote{Corresponding Author}
\affiliation{%
  \institution{Korea University}
  \city{Seoul}
  \country{Republic of Korea}}
\email{Sungho_Suh@Korea.ac.kr}

\renewcommand{\shortauthors}{Ray et al.}

\begin{abstract}
Prolonged seated activity is increasingly common in modern environments, raising concerns around musculoskeletal health, ergonomics, and the design of responsive interactive systems.
Existing posture sensing methods such as vision-based or wearable approaches face limitations including occlusion, privacy concerns, user discomfort, and restricted deployment flexibility.
We introduce ChairPose, the first full body, wearable free seated pose estimation system that relies solely on pressure sensing and operates independently of chair geometry.
ChairPose employs a two stage generative model trained on pressure maps captured from a thin, chair agnostic sensing mattress. Unlike prior approaches, our method explicitly incorporates chair morphology into the inference process, enabling accurate, occlusion free, and privacy preserving pose estimation.
To support generalization across diverse users and chairs, we introduce a physics driven data augmentation pipeline that simulates realistic variations in posture and seating conditions.
Evaluated across eight users and four distinct chairs, ChairPose achieves a mean per joint position error of 89.4 mm when both the user and the chair are unseen, demonstrating robust generalization to novel real world generalizability.
ChairPose expands the design space for posture aware interactive systems, with potential applications in ergonomics, healthcare, and adaptive user interfaces.
\revise{All code and data are publicly available on Kaggle at \href{https://www.kaggle.com/datasets/lalaray/chairpose}{ChairPose}.}
\end{abstract}


 \begin{teaserfigure}
 \centering
   \includegraphics[width=0.9\textwidth]{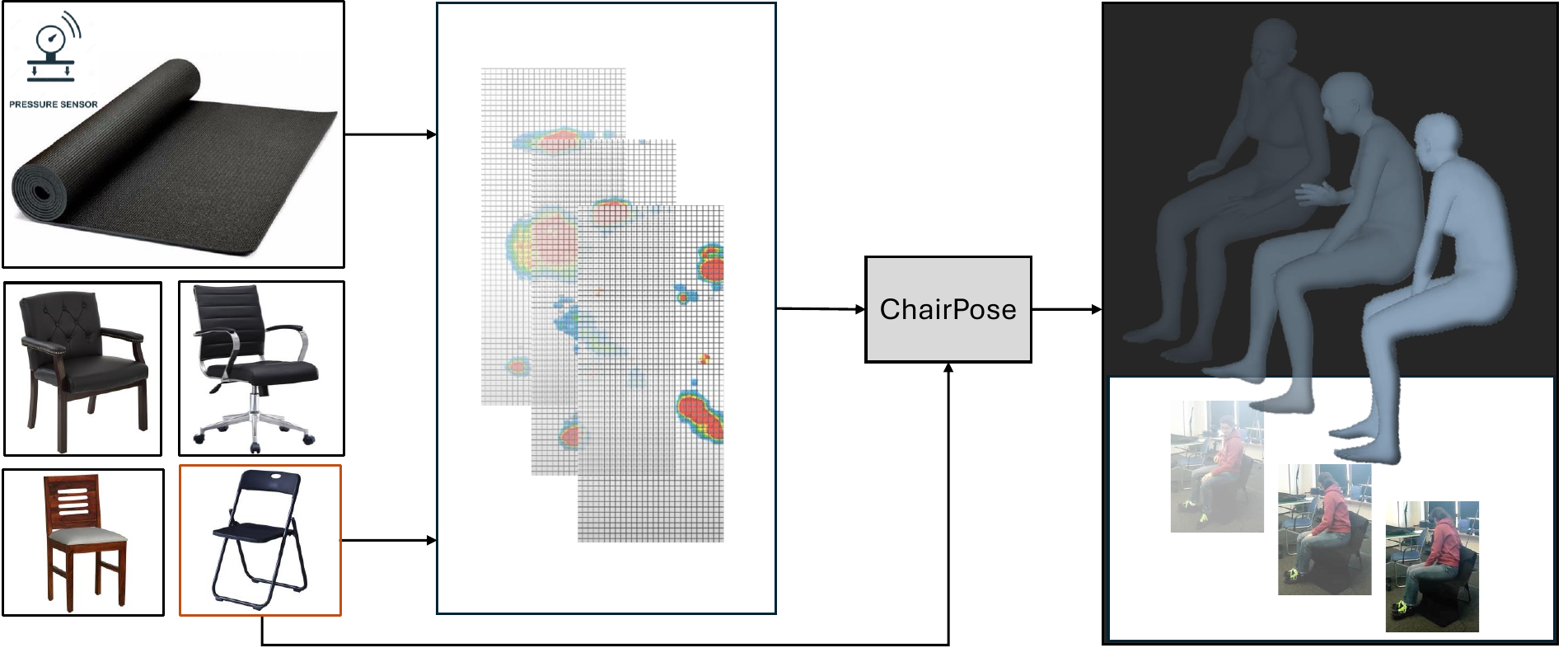}
   \Description{Diagram showing a pipeline where pressure sensor inputs and 3D scans of different chairs are processed by ChairPose to output full-body 3D pose estimations. Output visuals include 3D body meshes sitting in chairs and corresponding real-world images.}
   \caption{ChairPose predicts full-body volumetric pose sequence from pressure inputs while accounting for chair morphology without requiring retraining or finetuning across different chairs or users.}
   \label{fig:teaser}
 \end{teaserfigure}


\maketitle
\section{Introduction}
Chairs are ubiquitous fixtures in modern life, facilitating activities across work, leisure, mobility, and, increasingly, social interactions. 
Given that the average person spends over 9 hours per day seated \cite{wallmann2013sitting, mclaughlin2020worldwide}, prolonged sitting has become a major health concern, linked to musculoskeletal disorders, cardiovascular disease, and reduced cognitive performance. 
This presents a compelling opportunity for intelligent seating systems that monitor and improve seated behavior, enhancing both health and productivity through real-time posture-aware interventions.

Accurate seated posture assessment is critical across multiple domains, such as ergonomics \cite{vermander2024intelligent}, healthcare \cite{wang2024double}, and human-computer interaction \cite{zhong2024accurate}. 
Proper sitting posture plays a significant role in preventing musculoskeletal disorders \cite{celik2018determination, daneshmandi2017adverse}, improving comfort \cite{roynarin2024postural}, and promoting overall well-being \cite{gwak2024physiological}. 
For individuals with limited mobility who rely on chairs or wheelchairs for much of their daily lives, maintaining optimal posture is crucial to preventing pressure sores, managing pain, and preserving independence. 
Smart seating systems that understand the postures beyond the contact pressure distributions could offer personalized support, early detection of postural imbalances, and enhanced quality of life.

Existing approaches often use wearable sensors or vision-based systems, but both face limitations—wearables are intrusive and uncomfortable, while cameras raise privacy concerns and suffer from occlusion. ChairPose addresses this need with a novel, passive approach that reconstructs full-body seated pose using only pressure input, enabling unobtrusive, real-time posture monitoring without visual sensing or body-worn devices. 

Traditional seated pose estimation methods typically rely on vision-based techniques, such as cameras \cite{nadeem2024sitting, huang2024scalable} and depth sensors \cite{jin2025sitpose}. 
However, these approaches face limitations, including occlusion issues, privacy concerns, and reliance on environmental lighting conditions and camera placement.  
Non-visual sensor-based pose estimation technologies have emerged as promising alternatives, enabling continuous and unobtrusive seated posture tracking.
Existing sensor-based methods often use wearable sensors like inertial measurement units (IMUs) to track body posture \cite{nouriani2024vector, zhang2024dynamic, chen2025skeleton}. 
Although these sensors provide detailed motion data, they come with their own set of challenges. 
IMUs require precise placement on multiple body segments \cite{ramani2024imuoptimize}, which can cause discomfort and misalignment during prolonged use. 
Hence, they are impractical for continuous and non-intrusive posture monitoring in everyday seating scenarios.
Wearable capacitive sensors\cite{zhou2023mocapose, yu2024seampose, shimada2020physcap} offer a looser and more comfortable fit but can only estimate body relative pose and, like IMUs, still suffer from the limitations of being wearable, requiring consistent contact and calibration.

More recently, pressure-based pose estimation methods have gained attention for their privacy-friendly, non-intrusive, and environmentally robust nature, making them well suited for continuous real-time posture monitoring \cite{chen2024cavatar, hansmart, seong2024intelligent, zhao20243d}. 
Unlike wearable sensors that require direct attachment to the body - often causing discomfort or requiring precise placement - pressure sensors can be seamlessly integrated into seating surfaces without disrupting user comfort. 
In particular, fabric-type pressure sensors have proven effective in capturing fine-grained spatial pressure distribution across contact areas. 
These sensors can be leveraged to reconstruct detailed seated human poses, capturing nuances in weight distribution and seating behavior that wearable approaches may overlook.

\revise{Most existing pressure-based pose estimation systems are designed for flat, planar surfaces such as carpets \cite{9577856} or beds \cite{wu2024seeing} and fail to generalize across non-planar or irregular chair shapes. 
Recent developments introduced fabric-based sensor embedded directly into chairs to estimate full 3D sitting poses \cite{seong2024intelligent, zhao20243d}. However, real-world environments feature numerous chair forms, including office chairs, recliners, wheelchairs, and lounge chairs, each influencing posture differently. }

\begin{table*}[h]
\centering
\revise{
\caption{Comparison of pressure-based sitting pose estimation systems. Sensor types include force-sensitive resistors (FSR), analog tactile pressure sensors (ATPS), strain gauge type force sensor (SGFS), Piezoresistive pressure sensors (PPS), thin flexible pressure sensors (TFPS,) and thermoplastic polyolefins (TPE).}
}
\label{tab:related_work_comparison}
\begin{tabular}{ccccccc}
\toprule
\textbf{System} & \textbf{Sensor Type} & \textbf{Placement} & \textbf{Pose Modeling} & \textbf{Cross-surface generalization} & \textbf{Real-time} & \textbf{Temporal}\\
\midrule
Mutlu et al.\cite{mutlu2007robust} & Tekscan & Embedded & - & - & - & n/a\\
Bibbo et al.~\cite{bibbo2019sitting} & ATPS & Embedded & - & - & - & n/a\\
SenseChair~\cite{nishimura2023detection} & SGFS & Embedded & - & - & \checkmark & n/a\\
Mizumoto et al.\cite{mizumoto2020design} & Accelometer & Embedded & - & - & \checkmark & n/a\\
Aminosharieh et al.\cite{aminosharieh2022development} & FSR & Embedded & - & - & \checkmark & n/a\\
SPRS~\cite{tsai2023automated} & FSR & Embedded & - & - & \checkmark & n/a \\
IS~\cite{seong2024intelligent} & PPS  & Embedded & \checkmark & - & -  & - \\
IC~\cite{9577856} & TFPS & Embedded & \checkmark & n/a (Only planar surfaces) & \checkmark & \checkmark \\
3DHPE~\cite{zhao20243d} & TFPS & Embedded & \checkmark & - & -  & - \\
Wu et al.~\cite{wu2024seeing} & TFPS & External & \checkmark & n/a (Only planar surfaces) & - & \checkmark \\
\hline
\textbf{ChairPose (Ours)} & TPE & External & \checkmark & \checkmark & \checkmark & \checkmark \\
\bottomrule
\end{tabular}
\Description{This table compares various pressure-based sitting pose estimation systems across multiple criteria. Each row represents a different system from prior work, and the columns describe key characteristics of each approach. The final row highlights ChairPose , our proposed method, which uses an external TPE-based pressure mat, supports full-body 3D pose modeling, generalizes across diverse chair shapes, operates in real time, and models temporal dynamics—capabilities that most prior systems lack.}
\end{table*}

\revise{Developing a robust seated pose estimation model that adapts to these variations is essential for reliable posture monitoring.
Although it is possible to integrate pressure sensor arrays directly into chairs to estimate seated posture, as done with prior works, this requires hardware customization and model retraining for each specific chair type, which is time-consuming and labor-intensive. } 

\revise{To address these limitations, ChairPose introduces a a fully passive, pressure-only system that uses a flexible fabric-based sensor mattress that can wrap around any chair shape, accurately estimating posture without chair-specific calibration or retraining.
It avoids the discomfort and setup complexity of wearable sensors by relying solely on passive pressure sensing embedded in seating surfaces.}
The proposed ChairPose visualized in \cref{fig:teaser} generalizes across a wide range of chair types and shapes by leveraging a large-scale simulated dataset that captures diverse seated pressure distributions from users with varying body types and across different chair geometries. 
By training on this comprehensive and varied dataset, the model learns to recognize and adapt to subtle posture variations influenced by chair contours and support structures. 
It enables ChairPose to accurately estimate seated poses without requiring per-chair retraining or calibration, making it robust and deployable across real-world seating environments.

The main contributions of this work are as follows:

\begin{enumerate} 
\item \textbf{ChairPose}: A chair-agnostic, wearable-free, pressure-based full-body pose estimation system that incorporates chair morphology into seated pose prediction. Unlike most prior pressure approaches, ChairPose performs continuous 3D pose \emph{regression}, rather than pose \emph{classification}, estimating temporally coherent full-body postures directly from pressure input.
\item \textbf{TDSD} (Temporal Dynamic Sitting Dataset): A new dataset of temporally synchronized seated postures, pressure maps, and 3D chair scans containing 12 unique activities, 4 different chairs and 8 different participants.
\item \textbf{Physics-driven Data Augmentation}: A novel simulation pipeline to generate diverse pressure-pose pairs from existing mocap datasets and 3D chair models using ragdoll dynamics. 
\item \textbf{Comprehensive Evaluation}: Extensive quantitative and qualitative evaluation of ChairPose, with application demonstrations in posture monitoring, body center-of-mass estimation, and activity recognition.
 
\end{enumerate}

\section{Related Work}
\subsection{Sitting Pose Monitoring}
\label{realted}

\revise{Sitting pose monitoring has been extensively studied using a variety of sensing modalities, each offering different trade-offs in terms of accuracy, deployment complexity, user comfort, and scalability~\cite{krauter2024sitting}. Among these, pressure sensors particularly thin and flexible pressure sensors (TFPS) have emerged as the most common hardware for posture recognition~\cite{ashruf2002thin}. These sensors are often embedded in chairs or integrated into portable pads, enabling unobtrusive, long-term monitoring with minimal user awareness.}

\revise{Early systems focused on coarse posture classification, labeling discrete categories such as “upright,” “leaning,” or “slouched”~\cite{jin2025sitpose, li2023abnormal, mutlu2007robust, bibbo2019sitting, nishimura2023detection, aminosharieh2022development, tsai2023automated}. While computationally lightweight, these methods lack the spatial granularity and continuity required for applications in clinical or assistive contexts.}

\revise{Vision-based systems offer richer outputs by estimating 2D/3D joint positions or full-body meshes from RGB or depth images~\cite{zhu2023motionbert, shin2024wham}. However, their performance can degrade under occlusion, variable lighting, and viewpoint changes. Moreover, their use in real-world settings is constrained by privacy concerns, especially in offices or healthcare environments.}

\revise{Wearable systems using inertial measurement units (IMUs) or gyroscopes can capture detailed motion information~\cite{greenwood2021cnn, mollyn2023imuposer, wu2024soleposer, mizumoto2020design}, but their dependence on correct placement, user compliance, and potential discomfort limits their feasibility for passive, everyday monitoring.}

\revise{Other sensing modalities—such as RFID~\cite{feng2019you, feng2020sitr}, ultrasonic sensors~\cite{alattas2014detecting}, deformation sensors~\cite{ardito2021low, demmans2007posture}, and hybrid approaches—have also been explored. Each offers a different balance of comfort, accuracy, and deployability, but none fully satisfy the ideal of shape-agnostic, non-intrusive, and scalable posture monitoring.}

\revise{Pressure-based systems strike an appealing balance between accuracy, privacy, comfort, and cost. TFPS—based on piezoresistive, piezocapacitive, or piezoelectric materials—can be seamlessly integrated into seating surfaces with minimal user disruption and no need for calibration.
Existing pressure-based posture systems can be broadly grouped into:}

\revise{Embedded systems: These integrate sensors directly into the chair ~\cite{seong2024intelligent, zhao20243d}, offering reasonable accuracy but limited generalization due to tight coupling with specific chair geometries. Adapting to new chairs typically requires full retraining.}

\revise{External systems: Pressure mats placed over chairs, beds, or floors offer portability and ease of deployment~\cite{wu2024seeing, seong2024intelligent}. However, many are optimized for flat surfaces and underperform on the curved, non-planar geometries of everyday seating.}

\revise{ChairPose, in contrast, introduces a thin, flexible sensing mattress capable of performing full-body 3D pose reconstruction from pressure maps. Crucially, it incorporates explicit 3D chair geometry through scans, enabling generalization across different chair shapes without retraining or chair-specific calibration. ChairPose also leverages physics-based simulation and motion quantization to enhance robustness and estimation accuracy.}

\revise{\cref{tab:related_work_comparison} summarizes key differences among recent pressure-based sitting posture systems, comparing sensing type, placement, real-time capability, pose modeling capacity, and generalization performance.}

\begin{figure*}[!t]
\centering
\includegraphics[width=0.9\textwidth]{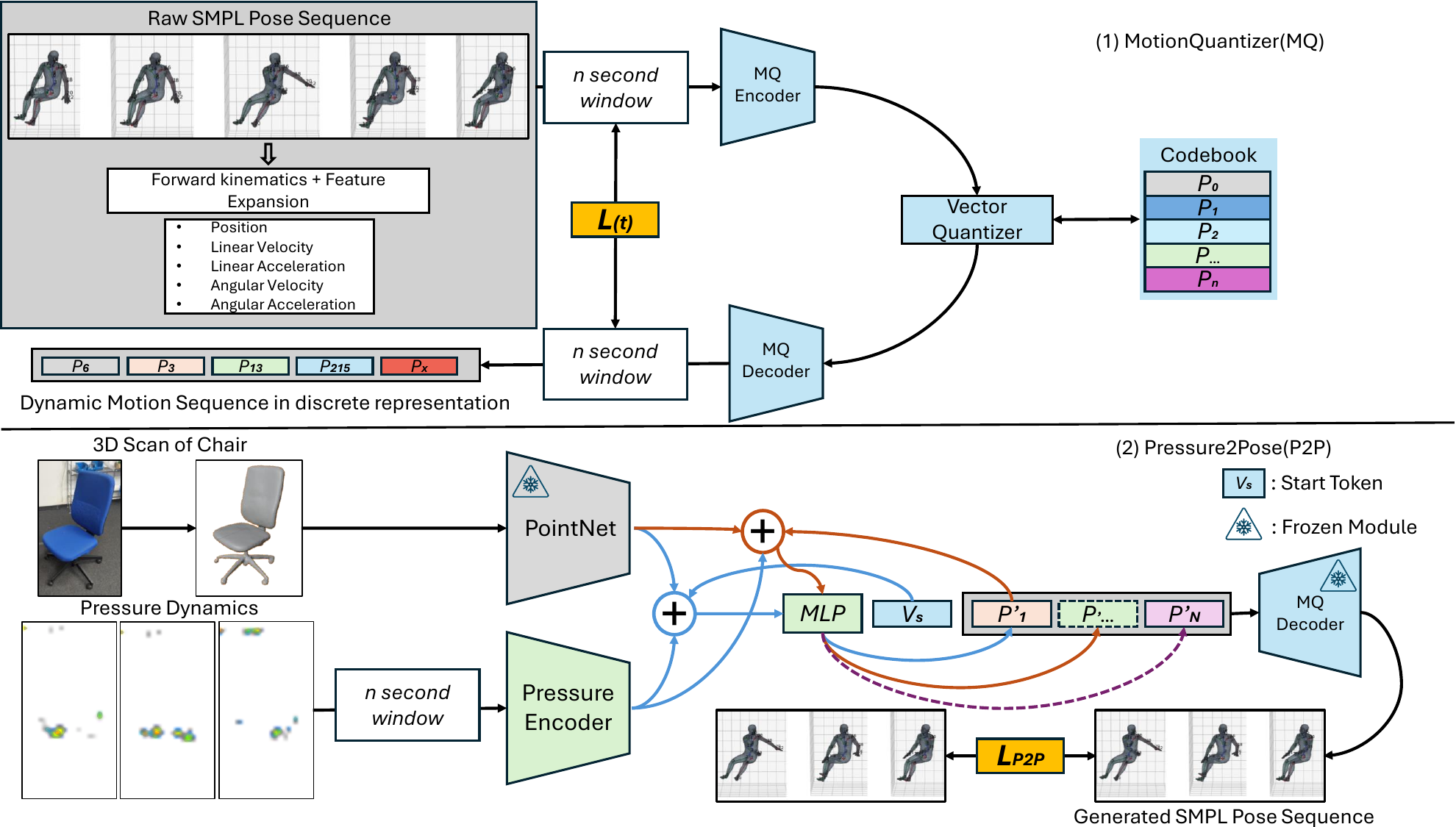} 
\Description{Two-part architecture diagram showing ChairPose’s training pipeline. The first module, MotionQuantizer, converts SMPL pose sequences into discrete motion tokens using a vector quantizer. The second module, Pressure2Pose, takes pressure inputs and 3D chair scans, processes them through encoders and a auto-regressive structure, and outputs predicted motion tokens that are decoded into full-body pose sequences.}
\caption{
ChairPose employs a two-step training architecture: first, the MQ module quantizes SMPL pose sequences into discrete motion tokens (P), capturing high-level temporal structure. Then, the P2P module learns to generate predicted motion tokens (P') directly from pressure dynamics. These predicted tokens are subsequently mapped back to continuous SMPL poses, enabling full-body reconstruction.} 
\label{fig:chairpose}
\end{figure*}

\subsection{Motion Tokenization}
With the rise of cross-modal large language models (LLMs) \cite{li2024llama, 10890689}, quantization techniques have gained popularity for transforming continuous time-series signals into discrete token sequences. These discrete representations not only preserve the core dynamics of the original signal but also enable the reconstruction of the signal from the tokenized vector sequence.

Given that 3D human motion is inherently a time-series signal, motion quantization—often using models such as Vector Quantized Variational Autoencoders (VQ-VAE) \cite{li2024semantically, goswami2024hypervq} has become a common approach in various generative tasks. These include tasks like text-to-motion generation \cite{fan2024freemotion, chen2024sato} and music-to-dance synthesis \cite{siyao2022bailando, zhuang2023gtn}. The primary advantage of quantizing continuous signals into a discrete latent space is that it significantly simplifies the learning process, allowing models to effectively capture complex temporal dependencies that would be challenging to learn directly from raw signals.

ChairPose builds on this idea with a two-stage generative framework: first quantizing motion into discrete tokens, and then learning to predict these tokens directly from pressure dynamics, enabling robust and interpretable seated pose reconstruction.

\subsection{Synthetic Pressure Map Generation}
Due to the lack of extensive real-world pressure sensor datasets, various synthetic pressure data generation techniques have been proposed to produce large-scale datasets for training pressure-based downstream models. 
One such model, PressNet \cite{scott2020image}, employs a neural network to generate pressure sensor data from 3D kinematic poses. 
However, as a supervised learning model, its ability is limited to generating pressure maps for postures already seen during training.

To overcome these limitations, recent approaches such as BodyPressureNet \cite{clever2022bodypressure} and PresSim \cite{ray2023pressim} combine physics-based 3D simulation with neural networks to generate pressure maps from depth images or 3D volumetric poses. 
Due to its architecture, BodyPressureNet is constrained to sleep postures. PresSim, the current state-of-the-art(SOTA) pressure simulation model, can generate pressure data across a wide range of body postures. Furthermore, it can operate directly on video input, making the system more practical and accessible.

ChairPose extends this line of work by adapting PresSim to simulate pressure dynamics on non-planar surfaces, an unaddressed gap in existing literature. This enables the generation of diverse pressure distributions across a wide range of seating environments, forming the basis for robust and generalizable pose estimation.

\section{ChairPose}
Seated posture recognition is essential for interactive systems, health monitoring, and accessibility applications. Yet, estimating full-body 3D pose from pressure input alone, particularly on arbitrary chairs, remains challenging. ChairPose addresses this by combining discrete motion representation with pressure-conditioned generative modeling, enabling robust, generalizable, and high-fidelity seated pose estimation, as visualized in \cref{fig:chairpose}. 
It consists of two key components: \textbf{MotionQuantizer} and \textbf{Pressure2Pose}, which transform a time-series pressure map sequence grounded with chair morphology into an accurate 3D volumetric pose sequence. 
The architecture employs a learned \textbf{codebook} to quantize full-body 3D pose sequences, effectively converting the regression problem into a classification task and significantly simplifying the learning process.

\subsection{MotionQuantizer}  
MotionQuantizer (MQ) is the core component of ChairPose, designed to encode full-body 3D motion into a compact, structured representation. 
Initially, a full-length motion sequence is segmented into fixed-size chunks spanning \(n\) seconds, each containing 15 frames (\(T = 15\)), 22 joints (\(J = 22\)), and 3 spatial dimensions (\(D = 3\)). 
This results in input tensors of size \((n \times T) \times J \times D\), which serve as the fundamental processing unit in MQ.

Within each chunk, MQ transforms continuous 3D pose data into discrete codes using a learned codebook, effectively converting the problem of pose estimation into a classification task. 
This discrete representation enables more efficient learning and improved generalization across users and seating conditions.

To enable effective quantization, MQ employs a Vector Quantized Variational Autoencoder (VQ-VAE) with a U-Net-based encoder-decoder architecture adapted from TMD \cite{ray2024text, ray2025txp}. 
The input to this model is derived from SMPL \cite{loper2023smpl} pose parameters (\(\theta\)), which are converted to 3D joint positions (\(\mathbf{P}\)) via forward kinematics. 
To capture dynamics beyond raw positions, we augment each chunk with motion descriptors:

\begin{itemize}
    \item Linear velocity: \( \mathbf{v}_l = \frac{d\mathbf{P}}{dt} \)
    \item Angular velocity: \( \mathbf{v}_a = \frac{d\theta}{dt} \)
    \item Linear acceleration: \( \mathbf{a}_l = \frac{d^2\mathbf{P}}{dt^2} \)
    \item Angular acceleration: \( \mathbf{a}_a = \frac{d^2\theta}{dt^2} \)
\end{itemize}

These motion cues enhance the richness of the representation, supporting more informative embeddings. The combined feature tensor is expressed as:  
\begin{equation}
    \mathbf{X} = [\theta, \mathbf{P}, \mathbf{v}_l, \mathbf{v}_a, \mathbf{a}_l, \mathbf{a}_a] \in \mathbb{R}^{(n \times T) \times J \times D}
\end{equation}

Training involves minimizing a combination of reconstruction loss (\(L_r\)) and quantization loss (\(L_q\)), weighted dynamically throughout training using scheduled annealing. This allows the model first to prioritize faithful reconstruction and later shift focus toward stable, efficient quantization:
\begin{equation}
    L(t) = w_r(t) \cdot L_r(t) + w_q(t) \cdot L_q(t)
\end{equation}

To further refine the codebook and prevent collapse, we incorporate two regularization strategies:
\begin{itemize}
    \item Exponential Moving Average (EMA): Smooths codebook updates to ensure stable usage of discrete tokens:  
    \begin{equation}
        \mathbf{C}^{(t)} = \alpha \mathbf{C}^{(t-1)} + (1 - \alpha) \mathbf{C}_{\text{new}}
    \end{equation}
    \item Quantization Dropout: Randomly drops quantized tokens during training to encourage robustness and improve generalization:  
    \begin{equation}
        \mathbf{q}_d = \mathbf{q} \cdot \mathbf{m}, \quad \mathbf{m} \sim \text{Bernoulli}(p)
    \end{equation}
    \end{itemize}

Through this architecture and training strategy, MQ compresses complex 3D motion into a discrete latent space while preserving temporal and structural integrity. The downstream P2P model then decodes these quantized representations to reconstruct realistic 3D human poses from pressure data.

\subsection{Pressure2Pose}
Pressure2Pose (P2P) is an autoregressive classifier that predicts full-body 3D human poses from pressure data collected during seated activities. At its core, P2P generates tokenized 3D poses one timestep at a time, conditioned on pressure input, chair geometry, and prior pose predictions—enabling temporally coherent motion synthesis.

The prediction process begins by segmenting the full input into discrete timesteps. At each timestep \( t \), the model takes in three inputs:

\begin{itemize}
    \item Pressure data \( \mathbf{P}_t \in \mathbb{R}^{N \times 80 \times 28} \): A sequence of pressure readings from \( N \) sensor frames, each with spatial resolution \( 80 \times 28 \) divided into n chunks based on the size of MQ quantization.
    \item Chair geometry \( \mathbf{M} \in \mathbb{R}^{5000 \times 3} \): A point cloud from a 3D mesh scan of the chair, processed through a pre-trained PointNet and feature extractor.
    \item Previous tokenized pose \( \mathbf{P}_{t-1} \in \mathbb{R}^{1 \times 512} \): The discrete pose vector from the prior timestep, predicted by the model itself.
\end{itemize}

At \( t = 0 \), this prior pose token is replaced by a learned start-of-sequence vector \( V_s \), initiating the autoregressive loop.

These three components are concatenated and passed through a linear layer to produce the current tokenized pose \( \mathbf{P'}_t \in \mathbb{R}^{1 \times 512} \). This output is then fed back as input at the next timestep, continuing the sequence generation until the entire series of poses \( \mathbf{P'}_1, \mathbf{P'}_2, \dots, \mathbf{P'}_N \) is produced.

Once the tokenized sequence is complete, each token \( \mathbf{P'}_t \) is passed through the pre-trained MQ Decode, which reconstructs the continuous 3D pose \( \hat{\mathbf{J}}_t \in \mathbb{R}^{J \times D} \), where \( J \) is the number of body joints and \( D \) is the spatial dimensionality. This results in the final predicted motion sequence $ \hat{\mathbf{J}}_1, \hat{\mathbf{J}}_2, \dots, \hat{\mathbf{J}}_N$.

P2P is trained using a combination of reconstruction and sequence-level objectives:

\begin{itemize}
    \item Reconstruction Loss: Since pose predictions are formulated as discrete token classification, a cross-entropy loss is applied between the predicted token and the ground-truth token \( \mathbf{P}_t^{\text{gt}} \):
    \begin{equation}
        L_{\text{reconstruction}} = -\sum_{i=1}^{C_n} \log \left( {P'}_{t,i} \cdot \mathbf{P}_{t,i}^{\text{gt}} \right)
    \end{equation}
    
    where \( C_n \) is the size of the codebook and \( \mathbf{P}_{t,i}^{\text{gt}} \) is the one-hot encoding of the target token.

    \item Sequence-Level Loss: To ensure temporal smoothness and global motion consistency, the entire sequence of decoded poses is compared to the ground truth using Mean Squared Error (MSE):
    \begin{equation}
        L_{\text{sequence}} = \frac{1}{N} \sum_{t=1}^{N} \| \hat{\mathbf{J}}_t - \mathbf{J}_t \|_2^2
    \end{equation}
\end{itemize}

The total loss combines both terms:
\begin{equation}
    L_{\text{P2P}} = L_{\text{reconstruction}} + \lambda L_{\text{sequence}}
\end{equation}

Where \( \lambda \) is a learnable hyperparameter that balances token accuracy and temporal consistency.

The autoregressive design of P2P enables continuous prediction of 3D human motion over time, even without prior motion history. Using the start token \( V_s \) ensures that generation can begin from the first frame. The model generates highly realistic and coherent pose sequences by fusing multimodal inputs—pressure signals, 3D chair geometry, and prior pose context.

Additionally, the discrete latent space learned by MQ simplifies pose prediction into a classification task, reducing complexity while preserving fidelity through the decoder.

\subsection{Baseline Method}
\label{baseline}
\revise{We adopt a simple but effective baseline model that directly regresses 3D human pose from raw pressure maps and chair geometry inputs, without employing motion quantization or temporal sequence modeling. This baseline is inspired by recent works MobilePoser \cite{xu2024mobileposer}and IMUPoser \cite{mollyn2023imuposer}, serving as a minimal benchmark to isolate the benefits of our proposed quantization and temporal modeling modules. }

\section{Data Synthesis and Training}
This section provides a comprehensive overview of our data pipeline, which includes real-world sensor data collection, the use of intelligent data augmentation strategies through the generation of synthetic samples using the PresSim framework \cite{ray2023pressim}, and the overall training setup employed to optimize our model. By combining real and simulated data, we significantly improve diversity, robustness, and generalizability in the training process.

\subsection{System Design}

\begin{figure}[!t]
\centering
\includegraphics[width=0.45\textwidth]{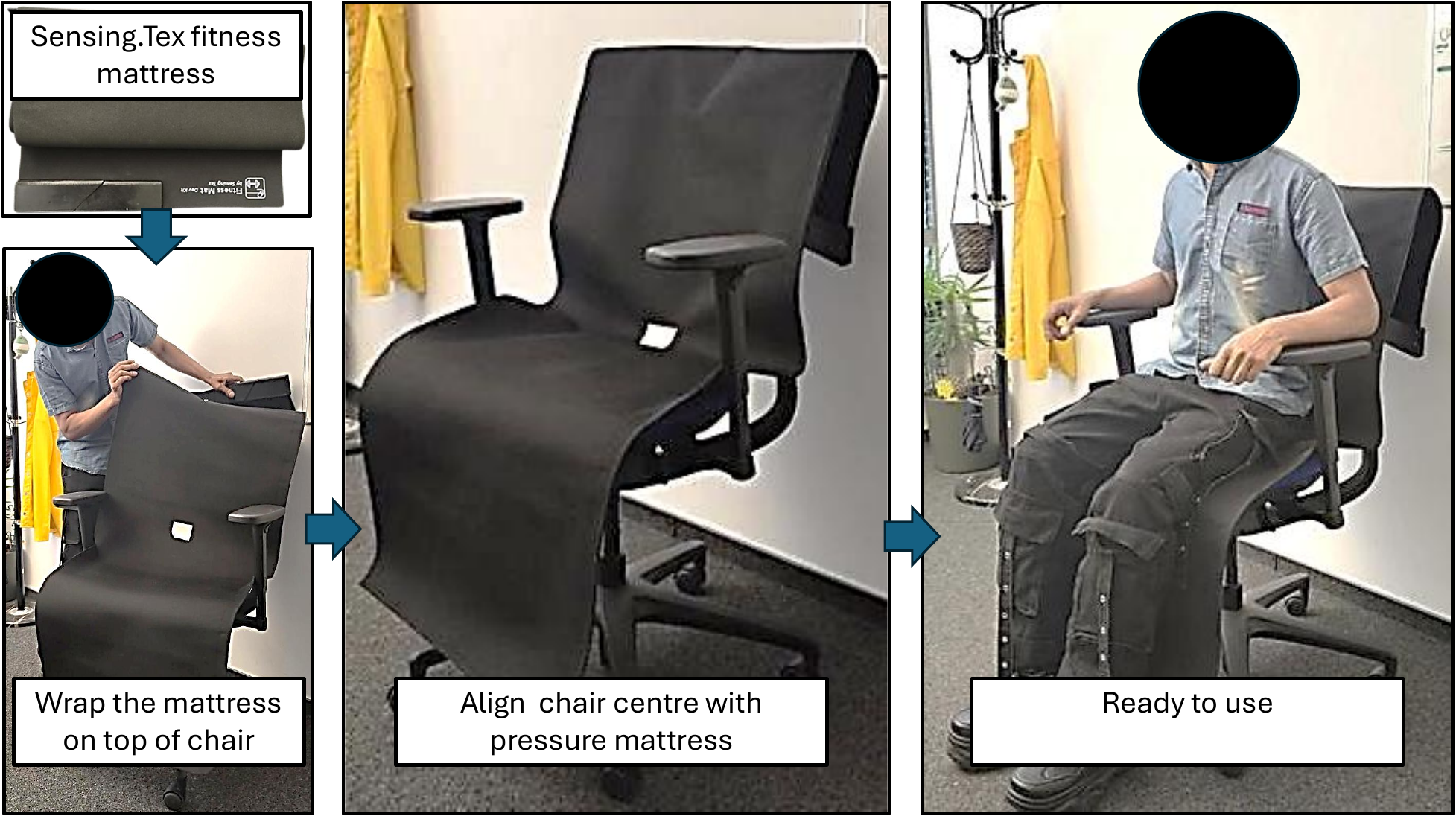}
\Description{Illustration showing how to set up the pressure sensor mat on a chair. The mat is aligned with the center of the chair and then placed on the seat. A person sits on the prepared chair, indicating the setup is complete and ready for use.}
\revise{\caption{Hardware setup: pressure mat placed on the chair surface. The flexible mat conforms to chair contours and requires no special calibration.}} 
\label{fig:ins}
\end{figure}

\revise{We introduce the Temporal Dynamic Sitting Dataset (TDSD), a multimodal dataset comprising synchronized pressure and 3D pose data from 8 participants across 4 chair types, covering 12 seated activities. The core sensing system consists of a flexible pressure mat and a vision-based depth capture device. As shown in \cref{fig:ins}, the pressure mat is simply placed over the chair surface, supporting fast deployment across diverse chair geometries without permanent modification.}

For pressure sensing, we used the Sensing.Tex Fitness Mat\footnote{\revise{Product website: }\url{https://www.sensingmat.cloud/}}, which consists of an 80$\times$28 sensor grid across a 560$\times$1680\,mm area. Each sensor measures 12$\times$16\,mm and records pressure within a 0–5000 mmHg range. The mat is made of non-slip, non-elastic thermoplastic polyolefins (TPE), allowing stable surface contact across various seating shapes and materials.

\revise{The 3D pose ground truth was captured using the Apple iPhone 16 Pro’s TrueDepth camera, positioned approximately 2 meters away at a 45° front-left angle to the seated subject. This fixed placement ensured consistent full-body coverage while minimizing occlusions. Meanwhile, the pressure mat was carefully aligned with the center of the seat and, when feasible, wrapped around the backrest to maximize contact area and accurately capture pressure distributions across the entire seating surface.}

Chair geometry was recorded using a Ferret Pro 3D scanner, enabling accurate reconstructions of the seating environment. These scanned meshes were later used for both physical interpretation and simulation-based augmentation.

\subsection{Data Acquisition and Synchronization}

\begin{figure}[!t]
\centering
\includegraphics[width=0.45\textwidth]{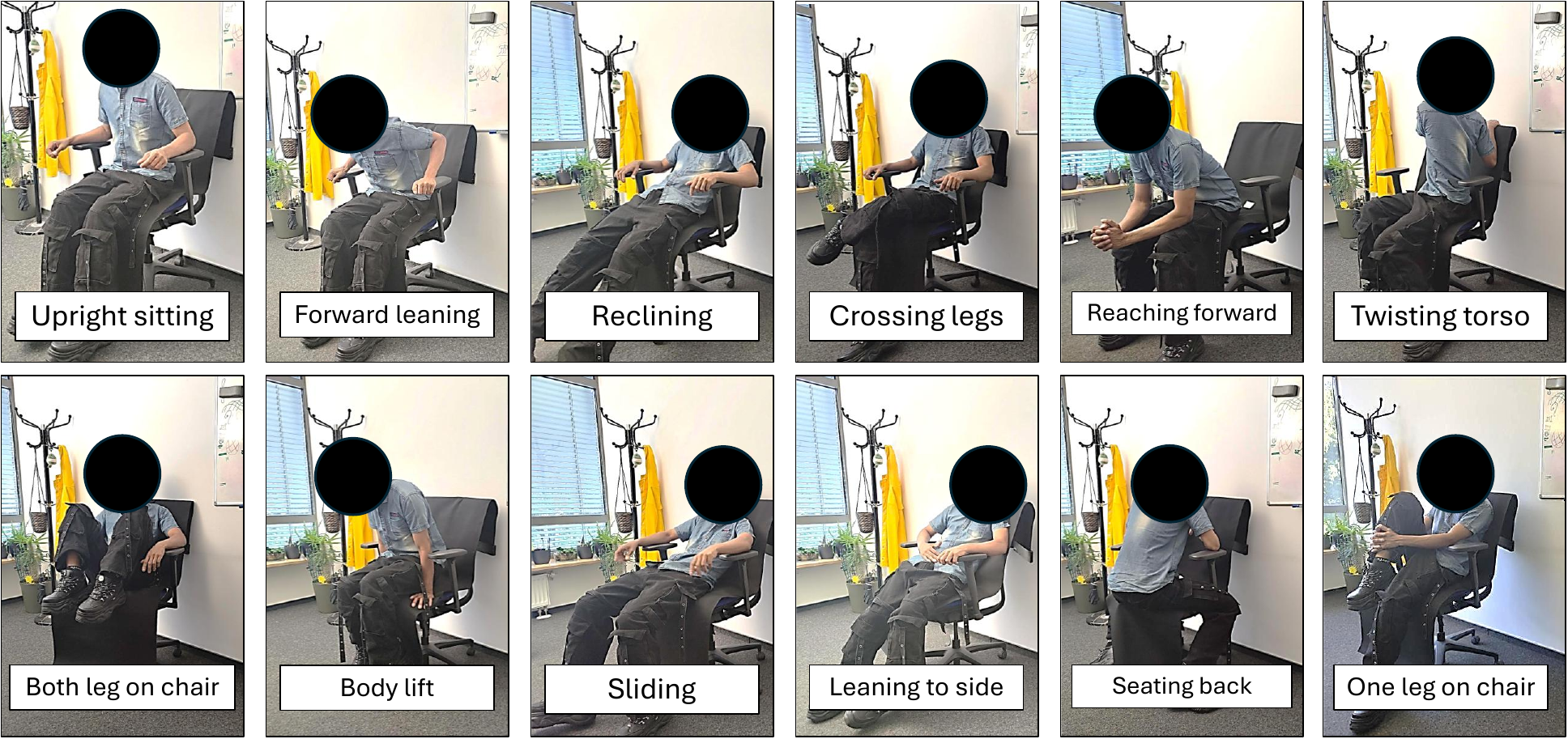}
\Description{Illustration showing 12 predefined seated actions performed by a person sitting on a chair. The actions include upright sitting, forward leaning, reclining, slouching, crossing legs, reaching forward, twisting torso, leg on chair, lifting body using hands, sliding on the chair, leaning to side, and sitting back.}
\revise{\caption{The 12 predefined seated actions used in data collection, showcasing diverse seated postures that capture a wide range of natural sitting behaviors for pose estimation.}}

\label{fig: data}
\end{figure}

\revise{Participants were asked to perform a set of 12 predefined seated actions given in \cref{fig: data}, reflecting common daily seated behaviors. These actions were displayed on-screen as short video prompts. Each participant repeated all actions 5 times per chair in a continuous sequence without breaks. This structure also allowed us to also leverage the transitions between actions as training data, capturing natural motion dynamics and posture shifts.}

\revise{All data collection sessions were conducted on four chair types with distinct designs: an office chair, a foldable chair, a barstool, and a manual wheelchair. This variation captures realistic differences in pressure distribution due to seat shape, backrest curvature, and support mechanics.}

\revise{To ensure synchronization between the pressure and pose streams, we used a simple but effective manual protocol. Participants tapped the seat surface three times at the beginning and end of each session. These taps were easily identifiable in both pressure and image streams and served as temporal anchors to align the sensor modalities.
}

\subsection{Post-processing}

Raw video sequences were processed using SMPLer-X \cite{cai2023smpler} to extract detailed 3D SMPL body mesh parameters. We chose this slower but higher-accuracy approach over traditional marker-based motion capture systems to reduce setup complexity while maintaining reliable pose fidelity. Our data collection setup followed best practices from the DMCB benchmark \cite{ray2023selecting, ray2024comprehensive}, which demonstrated that under controlled conditions and minimizing occlusion—monocular vision-based methods can achieve accuracy comparable to conventional MoCap systems.

The extracted pose sequences were filtered to remove outliers and ensure temporal smoothness and consistency. Corresponding pressure maps were downsampled and normalized to standardize input dimensions and ranges. All data streams were synchronized using a tap-based protocol to ensure precise temporal alignment. Each final data sample includes the SMPL pose parameters, the corresponding chair mesh identifier, and the time-aligned pressure map for each frame.

\begin{table}[!t]
\centering
\Description{Table listing basic statistics for 8 subjects in the TDSD dataset. Columns include Subject ID, Mass in kilograms, Height in centimeters, Gender, and total number of recorded frames. Subjects are a mix of male and female participants with varying body sizes. The total number of synchronized frames is 96,125.}
\caption{TDSD participant statistics and total synchronized frame count. Each subject performed 12 actions over 4 chair types, with 5 repetitions per action.}
\begin{tabular}{ccccc}
\hline
Subject & Mass (kg) & Height (cm) & Gender & Frames \\
\hline
1 & 65.3 & 175 & Male & 12,013 \\
2 & 70.9 & 175 & Male & 11,985 \\
3 & 65.4 & 176 & Female & 12,023 \\
4 & 54.9 & 162 & Female & 12,011 \\
5 & 60.1 & 157.5 & Female & 12,003 \\
6 & 57.8 & 162 & Male & 12,067 \\
7 & 51.0 & 160 & Female & 12,015 \\
8 & 62.4 & 172 & Male & 12,008 \\
\hline
\textbf{Total} & - & - & - & \textbf{96,125} \\
\hline
\end{tabular}

\label{tab:dataset_statistics}
\end{table}

\subsection{Synthetic Training Data Augmentation}

\begin{figure}[!t]
\centering
\includegraphics[width=0.45\textwidth]{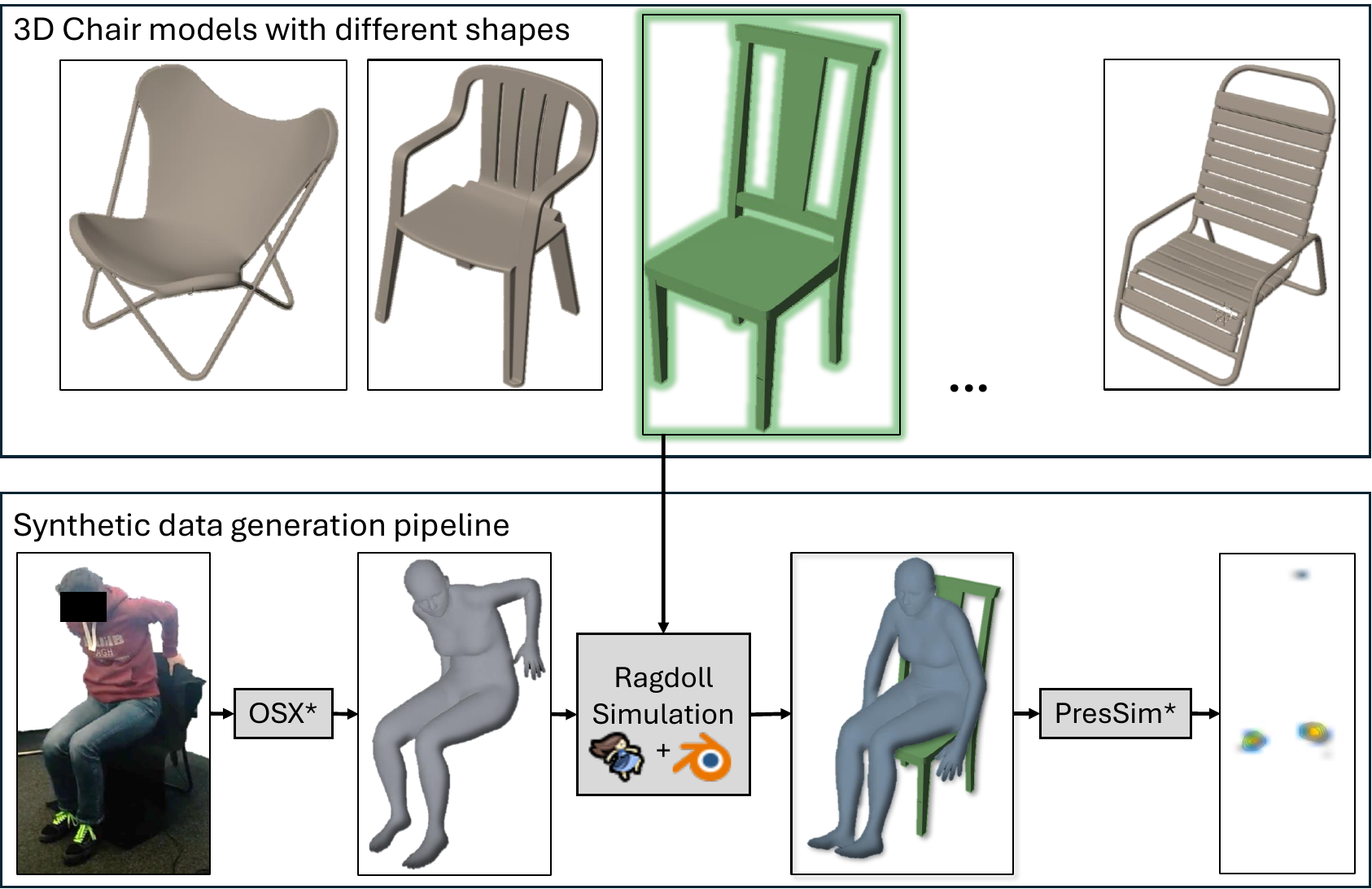} 
\Description{Diagram showing a data augmentation pipeline that uses various 3D chair models with different shapes. The pipeline combines motion capture data, ragdoll physics simulation, and a pretrained pressure sensor simulator to produce pressure data across diverse chair morphologies.}
\caption{Intelligent data augmentation pipeline that extends PresSim with ragdoll physics to generate diverse pressure data across varied chair morphologies from existing motion capture sequences.}
\label{fig:sim}
\end{figure}

Due to the limited real-world data variability for training the model and failure cases for capturing the complex physical interaction between the human body and various chair geometries by using simple data augmentation techniques (e.g., flipping, scaling, or noise injection), we adopt a more advanced strategy through intelligent data augmentation using the PresSim Framework.

PresSim extracts 3D human activity information from videos, simulates floor deformation, and produces realistic pressure sensor data. This significantly reduces the need for extensive real-world data collection in human activity recognition.

To tailor PresSim for our specific use case, we modified the floor deformation module by replacing the floor with 12 different CAD models of chairs sourced from the internet. Additionally, we integrated Ragdoll physics into the existing motion capture files using RagdollXBlender \cite{RagdollDynamics}. Ragdoll physics simulates realistic body dynamics by modeling the human body as a set of interconnected rigid bodies with physical constraints (e.g., joints, collisions, gravity, and mass distribution). By applying these simulations to motion capture sequences, we ensure that the character's interactions with each unique chair geometry obey plausible physical laws—such as falling into a chair, adjusting posture, or reacting to balance shifts.

This process results in chair-specific, physics-grounded motion capture data that better reflects how real bodies behave in various seated scenarios. These physically realistic motions generate more accurate deformation profiles, which are then processed through the PresSim pipeline to create a high-fidelity synthetic dataset for training our model. 
This, in turn, generates deformation profiles, which are processed through the PresSim pipeline to create a synthetic dataset for training our model.
The final synthetic data generated 12 more chair shapes with 1,153,500 frames of synchronized SMPL pose with pressure dynamics.

\subsection{Training Setup}

Our model is implemented in PyTorch and trained on a local Linux system equipped with an NVIDIA A6000 Ada Lovelace GPU and an AMD Ryzen 9 CPU. We used a combination of synthetic and real-world data for training, while evaluation was conducted exclusively on real data. \revise{The training process involved a batch size of 32, a learning rate of $1 \times 10^{-4}$, and the AdamW optimizer with a weight decay of $1 \times 10^{-5}$. Training was performed for a maximum of 200 epochs, with early stopping applied based on validation loss convergence (patience = 15 epochs). A cosine annealing learning rate scheduler was used to dynamically adjust the learning rate during training.}

\revise{To manage temporal dependencies and discrete motion tokenization, we employed a codebook size ($C$) of 1028 and a quantization window size ($n$) of 1 second. An exponential moving average (EMA) was applied to stabilize codebook updates, with a smoothing factor $\alpha = 0.99$. Additionally, Quantization Dropout was used during training with a dropout probability $p = 0.2$ to improve generalization. Sequence-level modeling incorporated a sequence loss weight $\lambda = 0.5$ to balance token accuracy and temporal consistency.}
\revise{Additional tasks such as activity classification (HAR), volumetric center of mass (VCoM) estimation, and posture monitoring were also performed on the same system.}

\section{Evaluation}
We evaluated our model using qualitative and quantitative methods, as well as through downstream use cases. To validate the accuracy of ChairPose, we employed leave-one-user-out cross-validation (LOUOCV), \revise{leave-one-chair-out cross-validation (LOCOCV) and leave-one-(chair+user)-out cross-validation (LOCUOCV) strategies}. Additionally, we conducted an ablation study to assess the contribution of each model component. For comparative evaluation, we benchmarked ChairPose against several state-of-the-art pressure-based methods—including IC \cite{9577856}, IS \cite{seong2024intelligent}, and 3DHPE \cite{zhao20243d}—as well as popular vision-based pose estimation systems such as SMPLer-X \cite{cai2023smpler} and MediaPipe \cite{singh2021real}, demonstrating ChairPose’s superior accuracy and robustness.

\subsection{Quantitative Evaluation}

\subsubsection{Metrics}
To evaluate the quantization by MQ module, we used FID$_k$, FID$_g$, and R-Precision from Bailando \cite{siyao2022bailando} and T2M \cite{guo2022generating}:
\begin{itemize}
    \item  FID$_k$ (Fréchet Inception Distance - kinetic (velocity, acceleration)) and FID$_g$ (Fréchet Inception Distance - geometric (position, angle)) measure the similarity between real and generated distributions:
  
  \begin{equation}
      \text{FID}(x, y) = \|\mu_x - \mu_y\|^2 + \mathrm{Tr}\left(\Sigma_x + \Sigma_y - 2(\Sigma_x \Sigma_y)^{1/2}\right)
  \end{equation}
  
  where \( \mu_x, \mu_y \) and \( \Sigma_x, \Sigma_y \) are the means and covariances of the real and generated features, respectively.

  \item R-Precision measures retrieval performance. For a generated sample, it checks if the correct real sample is among the top-k nearest real samples. It is computed as:

  \begin{equation}
    \text{R-Precision} = \frac{1}{N} \sum_{i=1}^N \mathbb{1}\{r_i \leq k\}
  \end{equation}

  where \( r_i \) is the rank of the truth of the ground among the samples recovered for the query \( i \) and \( \mathbb{1} \) is the indicator function.
\end{itemize}

To evaluate the accuracy of the estimated pose by the P2P module as compared to the ground truth pose, we used common metrics used for pose estimation models like MPJPE, PA-MPJPE, and MPVE, all in millimeters (mm):

\begin{itemize}
    \item MPJPE (Mean Per Joint Position Error) and PA-MPJPE (Procrustes Aligned MPJPE) used to evaluate joint-based pose: 
  \begin{equation}
    \text{MPJPE} = \frac{1}{J} \sum_{j=1}^{J} \| x - p_j \|_2
  \end{equation}

  where X is $\hat{p}_j$ for MPJPE and $R(\hat{p}_j))$ for PA-MPJPE with rigid (Procrustes) transformation $R$.
    \item MPVE (Mean Per Vertex Error): Used for evaluating the SMPL mesh-based predictions instead of joints:

  \begin{equation}
    \text{MPVE} = \frac{1}{V} \sum_{v=1}^{V} \| \hat{v}_v - v_v \|_2
  \end{equation}

  where \( V \) is the number of vertices, \( \hat{v}_v \) the predicted vertex, and \( v_v \) the ground truth.

\end{itemize}

\subsubsection{Results}

\paragraph{Ablation Study on Motion Quantization (MQ)}
\revise{We first conducted a comprehensive ablation study on the key Motion Quantization (MQ) component to identify optimal hyperparameters and verify the effectiveness of our refinement strategies.
\cref{tab:motion_quantizer_abl} summarizes experiments varying the quantization window size ($n$) and codebook size ($C$). Our results indicate that a 1-second window with a codebook size of 1028 achieves the best performance, yielding the lowest Fréchet Inception Distance (FID) scores — FID$_k$ = 58.18, FID$_g$ = 28.15 — and the highest R-Precision of 0.656, indicating a superior balance between reconstruction fidelity and diversity.
Next, \cref{tab:motion_quantizer_abl_2} presents a stepwise ablation on the best MQ configuration (1s window, 1028 codebook). Starting from the baseline MQ, incorporating the Feature Extractor (FE) module substantially lowers FID and improves R-Precision. Adding Exponential Moving Average (EMA) further enhances these metrics, and finally, applying Quantization Decoding (Q-D) yields the best overall performance. This confirms the cumulative benefits of these refinements in the MQ module.}

\paragraph{Ablation Study on Data Augmentation}
\revise{As a baseline comparison to our physics-driven synthetic data augmentation, we implemented a simpler, transformation-based data augmentation strategy applied directly to the real-world TDSD dataset. This baseline incorporates common geometric and sensor-level transformations, including horizontal flipping of pressure maps, random scaling of sensor values, and the addition of low-magnitude Gaussian noise to simulate minor variations in sensor readings.
}

\revise{We retrained the Pressure2Pose architecture under identical experimental conditions using only this baseline-augmented dataset, excluding any physics-simulated samples. This setup enables us to isolate and quantify the performance gains achieved through our intelligent, physics-grounded data generation pipeline compared to conventional data manipulation techniques. The evaluation results demonstrate that the baseline augmentation strategy underperforms not only in comparison to our simulation-assisted approach but also relative to models trained exclusively on real data. Specifically, MPJPE increased by approximately 42.8\% compared to the physics-based augmentation and by 2.3\% compared to training on real data alone. These findings underscore the limitations of simple augmentation methods in addressing domain shifts caused by user and chair variability, highlighting the importance of realistic physical modeling for achieving robust seated pose estimation across diverse users and seating environments.}

\paragraph{Baseline Performance and Component Contributions}
\revise{To isolate the impact of each core component in ChairPose, we introduced a baseline model that directly regresses 3D joint positions from raw pressure input, without Motion Quantization (MQ) or sequence-level modeling as described in \cref{baseline}. This baseline shares the same P2P backbone architecture as ChairPose, allowing a direct comparison of temporal discretization and autoregressive sequence modeling effects.
On LOUOCV, the baseline achieves 195.2 mm MPJPE, which MQ reduces to 56.7 mm, and the full model with sequence loss further improves to 53.5 mm. Similar trends occur on LOCOCV and LOCUOCV, where the baseline MPJPEs of 252.5 mm and 260.7 mm are reduced to 88.2 mm and 89.4 mm, respectively, by the full ChairPose. These results confirm that MQ and temporal modeling substantially boost accuracy beyond the baseline.
The baseline model's performance provides a clear reference point to demonstrate the improvements brought by MQ and temporal losses.}

\paragraph{Generalization Across Users and Chairs}
\revise{We evaluated ChairPose under three cross-validation protocols to rigorously test its ability to generalize across users, chairs, and their combinations:}
\begin{itemize}
    \item Leave-User-Out Cross-Validation (LOUOCV) results are in \cref{tab:mpjpe_results_louocv}.
    \item Leave-Chair-Out Cross-Validation (LOCOCV) results appear in \cref{tab:mpjpe_results_lococv}.
    \item \revise{Leave-(User+Chair)-Out Cross-Validation with broader domain shifts (LOCUOCV) is shown in \cref{tab:mpjpe_results_locuocv}}.
\end{itemize}

In the LOUOCV setting, ChairPose's full model (MQ + sequence loss) consistently outperforms its baseline and all adapted state-of-the-art (SOTA) pressure-based methods. It achieves the lowest errors MPJPE: 53.5 mm, PA-MPJPE: 37.9 mm, and MPVE: 44.2 mm confirming the effectiveness of motion discretization and temporal regularization for precise 3D pose estimation from pressure data.
In the more challenging LOCOCV protocol, even the baseline ChairPose model surpasses all existing pressure-based SOTA approaches (MPJPE 252.5 mm vs. IC 288.0 mm, 3DHPE 258.5 mm, and IS 265.2 mm). Incorporating MQ drastically reduces errors to 92.3 mm, with the full model further improving to 88.2 mm, highlighting strong generalization to novel chair geometries and motion dynamics.
\revise{Extending to LOCUOCV, which includes greater variability and domain shifts, ChairPose maintains robust performance. The full model achieves MPJPE of 89.4 mm, PA-MPJPE of 88.7 mm, and MPVE of 84.3 mm, demonstrating resilience to diverse users and furniture types.}

\paragraph{Impact of Data Augmentation Strategies}
\revise{To assess the effectiveness of our physics-driven synthetic data augmentation, we conducted an ablation study comparing it with a simpler, transformation-based baseline applied directly to the real-world TDSD dataset. This baseline involved common geometric and sensor-level manipulations, such as horizontal flipping of pressure maps, random scaling of sensor values, and low-magnitude Gaussian noise to simulate minor sensor variations.
We retrained the Pressure2Pose architecture using only this baseline-augmented dataset, excluding any simulation-generated samples, to isolate the value of physics-based augmentation. Results show that this naive augmentation strategy performs worse not only than our simulation-assisted approach but also compared to models trained solely on unaugmented real data. Specifically, MPJPE increased upto 85.5\% over physics-based augmentation and by 2.3\% over real-data-only training.}

\revise{In contrast, incorporating our simulation-assisted synthetic data led to substantial performance gains across all cross-validation settings. MPJPE was reduced from 71.8 mm to 53.5 mm in leave-one-user-out (LOUOCV), from 158.7 mm to 88.2 mm in leave-one-chair-out (LOCOCV), and from 164.2 mm to 89.4 mm in leave-one-user-and-chair-out (LOUCOCV)—corresponding to relative reductions of 25\%, 44\%, and 46\%, respectively. Similar trends were observed for PA-MPJPE and MPVE metrics.}

\revise{These results highlight that simple augmentation techniques are insufficient to address domain shifts caused by user and chair variability. In contrast, our physics-based simulation pipeline significantly improves generalization and robustness, confirming ChairPose’s ability to scale to unseen users and furniture without requiring additional manual data collection or calibration.}

\paragraph{Comparison to State-of-the-Art Vision and Pressure-Based Methods}
\revise{We adapted several leading SOTA methods to our experimental setup featuring non-planar pressure sensor arrays and diverse chair geometries to enable fair comparisons:}
\begin{itemize}
    \item IC \cite{9577856} is a temporal 3D pose estimator from planar pressure maps, modified to handle non-planar inputs.
    \item IS \cite{seong2024intelligent} accommodates non-planar chairs but lacks temporal modeling.
    \item 3DHPE \cite{zhao20243d} is a temporal 3D pose estimator assuming fixed chair geometry.
\end{itemize}
These methods lack chair-agnostic capabilities essential for our multi-chair evaluation. We re-implemented all methods in PyTorch with necessary modifications for fair benchmarking.

In addition, two vision-based pose estimation systems were evaluated:
\begin{itemize}
    \item SMPLer-X \cite{cai2023smpler}: Provides accurate volumetric meshes and serves as the source of ground-truth annotations for our dataset. However, it requires offline processing, taking approximately 5 seconds per frame, making it unsuitable for real-time or continuous monitoring applications.
    \item MediaPipe \cite{singh2021real}: Offers real-time skeletal pose estimates but with sparse keypoints and limited 2D reasoning, resulting in significantly lower accuracy.
\end{itemize}

\revise{Across all three evaluation protocols ChairPose consistently outperforms existing pressure-based state-of-the-art methods by a significant margin.
In LOUOCV, ChairPose (MQ + sequence loss) achieves an MPJPE of 53.5 mm, drastically improving over prior pressure-based methods like IC (220.5 mm), 3DHPE (168.4 mm), and IS (196.1 mm). Vision-based approaches such as SMPLer-X achieve notably lower errors (~11 mm) but require highly controlled lab environments with fixed lighting and camera setups.
Under the more challenging LOCOCV and LOCUOCV protocols—which introduce unseen chair shapes and unseen users plus chairs, respectively—ChairPose maintains superior performance among pressure-based models. For example, in LOCOCV, ChairPose reduces MPJPE to 88.2 mm compared to IC’s 288.0 mm and in LOCUOCV to 89.4 mm versus IC’s 298.7 mm. These results highlight ChairPose’s robustness to variations in seating geometry and subjects, narrowing the gap to vision-based methods while relying solely on pressure sensing.
MediaPipe provides faster, real-time estimation but suffers from sparse keypoints and higher errors (approx 47–52 mm across all tests), remaining less accurate than ChairPose’s full model.
Overall, ChairPose sets a new performance standard for pressure-based 3D pose estimation, significantly closing the accuracy gap to vision-based methods without their practical constraints, enabling deployment in privacy-sensitive environments and situations where cameras are impractical, occluded, or misaligned. It is inherently invariant to lighting, background clutter, occlusion, and camera viewpoint variations—factors that commonly degrade vision-based methods in real-world settings.
Operating passively with minimal power consumption and no visual data capture, ChairPose eliminates privacy concerns. It imposes no line-of-sight constraints and is robust against body self-occlusion (e.g., crossed arms), reflective surfaces, and clothing variations. These attributes make ChairPose especially well-suited for long-term, in-home monitoring, seated posture analysis, rehabilitation, and healthcare environments where privacy, energy efficiency, and robustness are critical.}

\begin{table}[!t]   
\Description{Ablation table evaluating different quantization window lengths (n) and codebook sizes (C) in the MQ module. Metrics reported are FID (lower is better), and R-Precision (higher is better). The best results are achieved with n = 1 second and C = 1028, showing the lowest FID, and highest R-Precision.}
\caption{Ablation study for MQ module: optimal quantization window length (n) and codebook size (C), evaluated by FID$_k$, FID$_g$, and R-Precision metrics.}
\begin{tabular}{ccccc}
\hline
n (sec) & C & FID$_k$ $\downarrow$ & FID$_g$ $\downarrow$ & R-Precision $\uparrow$ \\
\hline
1 & 256 & 75.20 $\pm$ 0.45 & 42.30 $\pm$ 0.38 & 0.512 $\pm$ 0.024 \\
2 & 256 & 77.50 $\pm$ 0.47 & 44.00 $\pm$ 0.40 & 0.498 $\pm$ 0.025 \\
5 & 256 & 80.10 $\pm$ 0.50 & 46.20 $\pm$ 0.42 & 0.483 $\pm$ 0.027 \\
1 & 512 & 65.40 $\pm$ 0.38 & 33.80 $\pm$ 0.30 & 0.602 $\pm$ 0.025 \\
2 & 512 & 67.80 $\pm$ 0.40 & 35.60 $\pm$ 0.32 & 0.588 $\pm$ 0.026 \\
5 & 512 & 70.10 $\pm$ 0.42 & 37.30 $\pm$ 0.34 & 0.573 $\pm$ 0.026 \\
1 & 1028 & \textbf{58.18} $\pm$ \textbf{0.32} & \textbf{28.15} $\pm$ \textbf{0.28} & \textbf{0.656} $\pm$ \textbf{0.026} \\
2 & 1028 & 60.00 $\pm$ 0.34 & 29.80 $\pm$ 0.29 & 0.642 $\pm$ 0.027 \\
5 & 1028 & 62.10 $\pm$ 0.36 & 31.20 $\pm$ 0.30 & 0.628 $\pm$ 0.027 \\
\hline
\end{tabular}
\label{tab:motion_quantizer_abl}
\end{table}

\begin{table}[!t]
\Description{Step-wise ablation study of MQ showing the impact of adding Feature Expansion (FE), Exponential Moving Average (EMA), and Quantization Dropout (Q-D). Metrics include FID (lower is better), and R-Precision (higher is better). Performance improves progressively with each component added. The full model, MQ + FE + EMA + Q-D, achieves the best results across all metrics.}

\caption{Step-wise ablation study demonstrating improvements from Feature Expansion (FE), Exponential Moving Average (EMA), and Quantization Dropout (Q-D) for MQ, evaluated by FID$_k$, FID$_g$, and R-Precision metrics.}
\begin{tabular}{lccccc}
\hline
Model &  FID$_k$ $\downarrow$ & FID$_g$ $\downarrow$ & R-Precision $\uparrow$ \\
\hline
MQ  & 82.50 $\pm$ 0.50 & 48.10 $\pm$ 0.42 & 0.460 $\pm$ 0.023 \\
MQ + FE  & 70.30 $\pm$ 0.44 & 38.90 $\pm$ 0.36 & 0.542 $\pm$ 0.025 \\
MQ + FE + EMA  & 61.90 $\pm$ 0.37 & 31.60 $\pm$ 0.30 & 0.630 $\pm$ 0.026 \\
\makecell[l]{MQ + FE + EMA \\ + Q-D}  & \textbf{58.18} $\pm$ \textbf{0.32} & \textbf{28.15} $\pm$ \textbf{0.28} & \textbf{0.656} $\pm$ \textbf{0.026} \\
\hline
\end{tabular}

\label{tab:motion_quantizer_abl_2}
\end{table}

\begin{table}[!t]
\Description{Comparison of ChairPose and baseline models against state-of-the-art pressure- and vision-based pose estimation methods on LOUOCV using MPJPE, PA-MPJPE, and MPVE metrics. ChairPose (MQ + L_sequence) achieves the best performance with the lowest errors across all three metrics: MPJPE 53.5, PA-MPJPE 37.9, and MPVE 44.2. Vision-based methods SmplerX and MediaPipe have lower errors than ChairPose (MQ + L_sequence). IC is noted to be adapted for non-planar input.}
\caption{MPJPE, PA-MPJPE, and MPVE results for LOUOCV comparing ChairPose with baseline and other SOTA  pressure and vision-based pose estimation models. $*$ is modified to support non-planar pressure input to have a fair comparison.}
\begin{tabular}{lccc}
\hline
Model & MPJPE$\downarrow$ & PA-MPJPE$\downarrow$ & MPVE$\downarrow$ \\
\hline
SmplerX \cite{cai2023smpler}  & 11.2$\pm$ 0.91 & 10.1$\pm$ 0.73 & 11.38$\pm$ 1.21 \\
MediaPipe \cite{singh2021real} & 47.43$\pm$ 0.58 & 38.1$\pm$ 0.77 & - \\
\hline
IC$^{*}$ \cite{9577856}  & 220.5$\pm$ 1.20 & 155.4$\pm$ 1.10 & - \\
3DHPE \cite{zhao20243d} & 168.4$\pm$ 0.85 & 112.5$\pm$ 0.90 & - \\
IS \cite{seong2024intelligent} & 196.1$\pm$ 0.88 & 171.0$\pm$ 0.89 & - \\
\hline
\makecell[l]{ChairPose\\ (Baseline)} & 195.2$\pm$ 0.37 & 169.7$\pm$ 0.74 & 181.4$\pm$ 0.27 \\
\makecell[l]{ChairPose\\ (MQ)} & 56.7$\pm$ 0.49 & 39.5$\pm$ 0.41 & 46.1$\pm$ 0.38 \\
\makecell[l]{ChairPose\\ (MQ + \\$L_{sequence}$)} & \textbf{53.5}$\pm$ \textbf{0.53} & \textbf{37.9}$\pm$ \textbf{0.26} & \textbf{44.2}$\pm$ \textbf{0.26} \\
\hline\end{tabular}

\label{tab:mpjpe_results_louocv}
\end{table}

\begin{table}[!t]
\caption{MPJPE, PA-MPJPE, and MPVE results for LOCOCV comparing ChairPose with baseline and other SOTA  pressure and vision-based pose estimation models. $+$ is modified to support dynamic chair shape input to have a fair comparison.}
\Description{Table comparing ChairPose with baseline and state-of-the-art vision and pressure-based models under the LOCOCV protocol using MPJPE, PA-MPJPE, and MPVE metrics. ChairPose with MQ and sequence loss achieves the best performance with MPJPE 88.2, PA-MPJPE 77.1, and MPVE 81.6. Other pressure-based models adapted for dynamic chair input, such as IC+, 3DHPE+, and IS+, show significantly higher errors. SMPLer-X achieves the lowest error overall, but ChairPose narrows the gap while supporting pressure-based sensing.}
\begin{tabular}{lccc}
\hline
Model & MPJPE$\downarrow$ & PA-MPJPE$\downarrow$ & MPVE$\downarrow$ \\
\hline
SMPLer-X \cite{cai2023smpler}  & 11.4$\pm$ 0.89 & 10.3$\pm$ 0.75 & 11.52$\pm$ 1.18 \\
MediaPipe \cite{singh2021real} & 47.60$\pm$ 0.61 & 38.3$\pm$ 0.79 & - \\
\hline
IC$^{+}$ \cite{9577856}  & 288.0$\pm$ 2.40 & 214.0$\pm$ 1.90 & - \\
3DHPE$^{+}$ \cite{zhao20243d} & 258.5$\pm$ 2.10 & 179.9$\pm$ 1.85 & - \\
IS$^{+}$ \cite{seong2024intelligent} & 265.2$\pm$ 2.34 & 185.7$\pm$ 1.95 & - \\
\hline
\makecell[l]{ChairPose\\ (Baseline)} & 252.5$\pm$ 2.21 & 178.5$\pm$ 1.86 & 192.5$\pm$ 1.75 \\
\makecell[l]{ChairPose\\ (MQ)} & 92.3$\pm$ 1.61 & 81.6$\pm$ 0.91 & 85.3$\pm$ 0.85 \\
\makecell[l]{ChairPose\\ (MQ + \\$L_{sequence}$)} & \textbf{88.2}$\pm$ \textbf{1.45} & \textbf{77.1}$\pm$ \textbf{0.88} & \textbf{81.6}$\pm$ \textbf{0.71} \\
\hline
\end{tabular}
\label{tab:mpjpe_results_lococv}
\end{table}

\begin{table}[!t]
\revise{\caption{MPJPE, PA-MPJPE, and MPVE results for Leave-(User+Chair)-Out Cross-Validation (LOCUOCV) comparing ChairPose with baseline and other state-of-the-art pressure- and vision-based pose estimation models. $+$ indicates modifications to support dynamic chair shape input for fair comparison.}}
\Description{Table comparing ChairPose with baseline and state-of-the-art vision and pressure-based models under the LOCUOCV protocol using MPJPE, PA-MPJPE, and MPVE metrics. ChairPose with MQ achieves MPJPE of 89.4 mm, PA-MPJPE of 88.7 mm, and MPVE of 84.3 mm, comparable to LOCOCV results. The full model (MQ + $L_{sequence}$) achieves the best performance, narrowing the gap with SMPLer-X while leveraging pressure-based sensing. Other pressure-based models adapted for dynamic chair input (IC+, 3DHPE+, IS+) show significantly higher errors.}
\begin{tabular}{lccc}
\hline
Model & MPJPE$\downarrow$ & PA-MPJPE$\downarrow$ & MPVE$\downarrow$ \\
\hline
SMPLer-X \cite{cai2023smpler}  & 11.2$\pm$ 1.02 & 10.9$\pm$ 0.85 & 11.7$\pm$ 1.35 \\
MediaPipe \cite{singh2021real} & 47.1$\pm$ 0.68 & 38.7$\pm$ 0.84 & - \\
\hline
IC$^{+}$ \cite{9577856}  & 298.7$\pm$ 2.65 & 222.3$\pm$ 2.05 & - \\
3DHPE$^{+}$ \cite{zhao20243d} & 267.8$\pm$ 2.30 & 187.6$\pm$ 2.00 & - \\
IS$^{+}$ \cite{seong2024intelligent} & 273.5$\pm$ 2.48 & 192.1$\pm$ 2.10 & - \\
\hline
\makecell[l]{ChairPose\\ (Baseline)} & 260.7$\pm$ 2.45 & 185.7$\pm$ 2.00 & 198.1$\pm$ 1.90 \\
\makecell[l]{ChairPose\\ (MQ)} & 92.3$\pm$ 1.61 & 85.0$\pm$ 0.91 & 88.7$\pm$ 0.85 \\
\makecell[l]{ChairPose\\ (MQ + \\$L_{sequence}$)} & \textbf{89.4}$\pm$ \textbf{1.61} & \textbf{88.7}$\pm$ \textbf{0.91} & \textbf{84.3}$\pm$ \textbf{0.85} \\
\hline
\end{tabular}
\label{tab:mpjpe_results_locuocv}
\end{table}

\begin{figure*}[!t]
\centering
\includegraphics[width=0.9\textwidth]{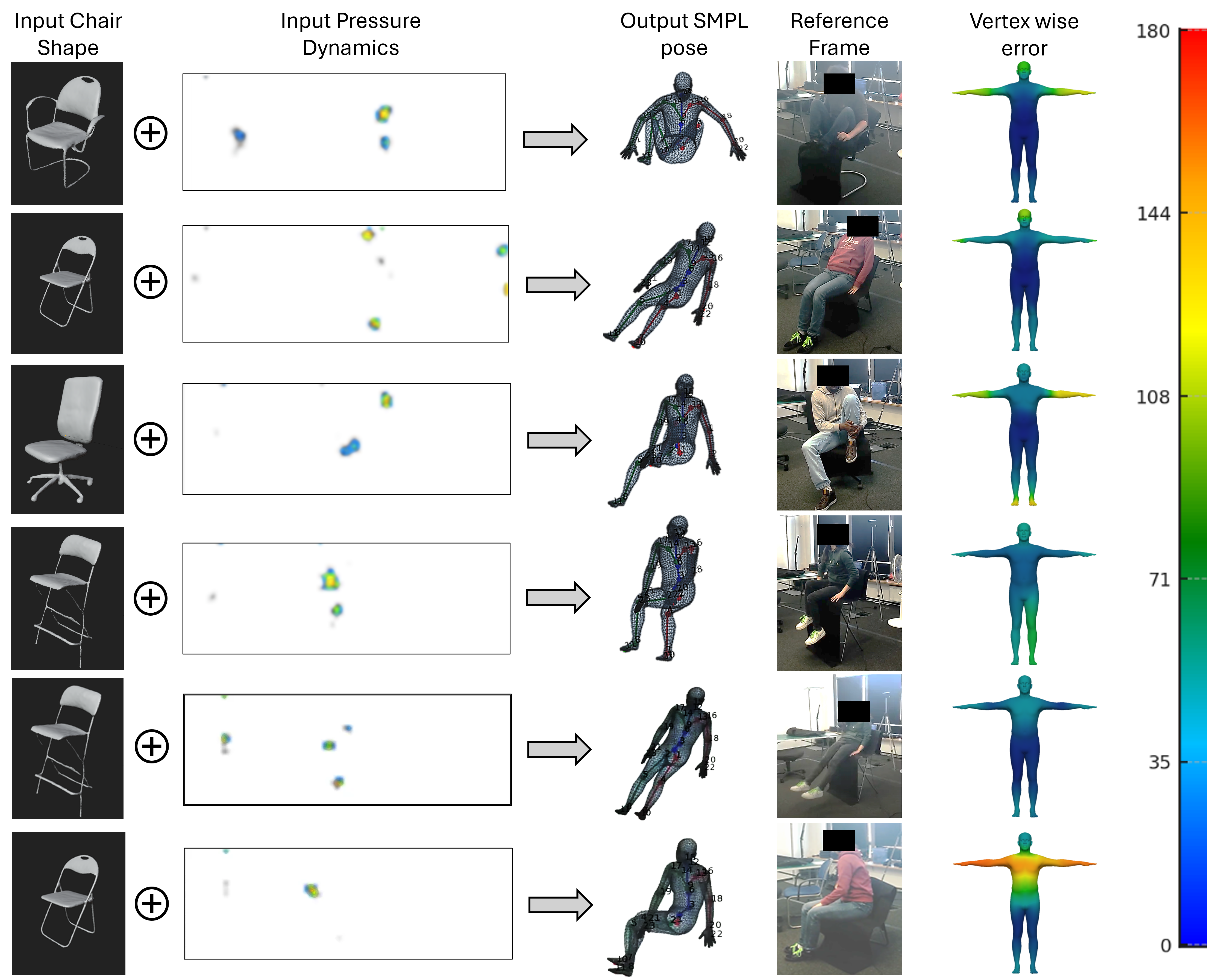} 
\Description{Grid visualization showing four examples of the ChairPose pipeline. Each row includes a 3D model of a different chair shape, its corresponding input pressure dynamics map, the resulting predicted SMPL body pose, and a real-world reference frame of a user in that chair. Demonstrates accurate pose predictions across varied chairs and users.}
\revise{\caption{
Visualization of the input pressure map, 3D chair point cloud, and predicted SMPL body pose, error color-map as compared to ground truth, shown alongside the reference video frame across different users and chair configurations.}} 
\label{fig:res}
\end{figure*}

\subsection{Qualitative Evaluation}

We present qualitative results, including visualizations and reference video captures in \cref{fig:res}, to illustrate ChairPose’s prediction performance. Interestingly, body parts such as the hands and head—which do not directly influence the pressure signal—still exhibit coherent and plausible motion. While not always perfectly accurate when compared to the reference video, these predictions maintain anatomical consistency and temporal smoothness. This behavior is largely enabled by the MQ module, which jointly quantizes all joints, allowing the P2P module to infer motion tokens that reflect the most likely full-body pose, even for joints lacking direct sensor input.

ChairPose’s robustness is further supported by data-driven training and the integration of 3D physics-based simulations for data augmentation. These elements help the model predict complete body poses even under partial sensor contact. This paradigm of using large-scale motion priors and physically grounded simulations aligns with recent work such as IMUPoser \cite{mollyn2023imuposer}, MoCaPose \cite{zhou2023mocapose}, SeamPose \cite{yu2024seampose}, and PhysCap \cite{shimada2020physcap}, which similarly use learned representations to infer full-body configurations from sparse or indirect inputs.

\section{Applications}
\begin{figure}[!t]
\centering
\includegraphics[width=0.45\textwidth]{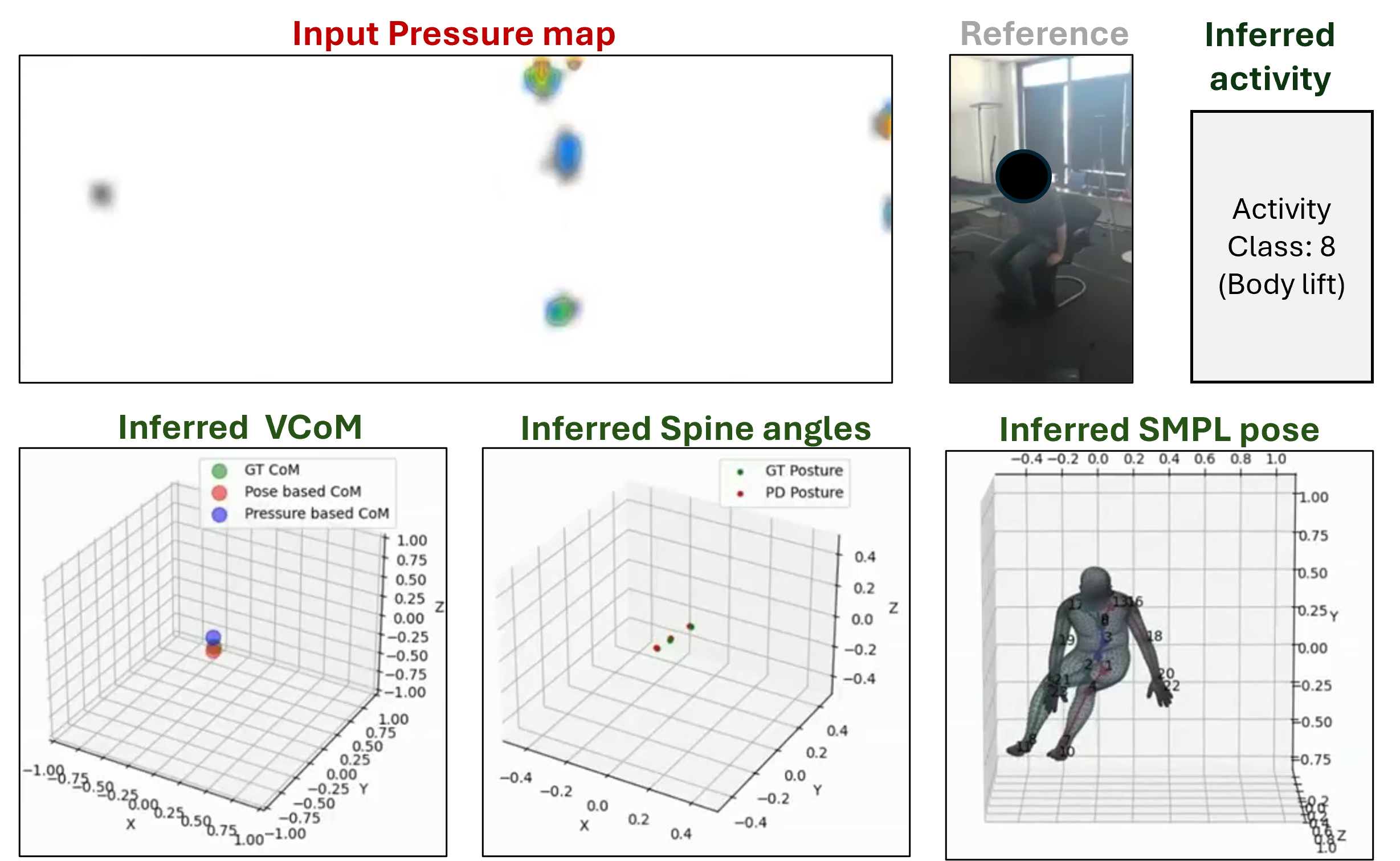} 
\Description{This figure demonstrates ChairPose’s ability to infer full-body seated pose from pressure input, along with derived metrics such as volumetric center of mass (VCoM) and spine angles. The top row shows the input pressure map and reference activity class (Body lift), while the bottom row illustrates the inferred VCoM, spine angles, and SMPL mesh-based pose. The results highlight ChairPose’s accuracy in reconstructing body posture and estimating key biomechanical features, showcasing its utility for applications such as posture monitoring, activity recognition, and ergonomic analysis.}
\revise{\caption{ChairPose's full body seated pose estimation and derived posture feedback, VCoM estimation, and enhanced activity recognition.}} 
\label{fig:abla}
\end{figure}

\revise{ChairPose enables full-body seated pose estimation from pressure input without requiring visual sensors or wearable devices. Based on this capability, we have implemented and demonstrated three core downstream tasks: posture feedback , volumetric center of mass (VCoM) estimation , and enhanced activity recognition. These are fully functional components built on top of ChairPose’s pose estimation pipeline as shown in \cref{fig:abla}.}

\revise{In addition to these implemented capabilities, we also illustrate conceptual use cases in automotive (driver drowsiness detection), workplace ergonomics (prolonged sitting feedback), and gaming (posture-based interaction) to highlight the broader potential of ChairPose. These scenarios are not implemented in the current system but are included to inspire future applications and deployments.}

\subsection{Body Center of Mass Estimation}

Accurate estimation of the body’s volumetric center of mass (VCoM) is critical for applications such as seated balance assessment, fall-risk prediction, and biomechanics research. Traditional pressure-based methods often attempt to directly regress the VCoM from sensor data using deep learning, but these approaches can struggle with generalization and anatomical consistency.

ChairPose takes a different approach by first reconstructing the full-body SMPL mesh from pressure dynamics. We then estimate the VCoM using SMPL anthropometric data \cite{yan2021learning}, computing a weighted average of the VCoM of individual body segments. This mesh-based method incorporates the full-body structure, allowing for more anatomically grounded and physically plausible predictions.
As shown in \cref{tab:com_results}, our approach achieves significantly lower Mean Absolute Error (MAE) and Root Mean Square Error (RMSE) compared to direct regression methods, aligning more closely with ground truth. These results highlight the benefit of full-body modeling for robust and accurate VCoM estimation in seated contexts.

\begin{table}[!t]
\Description{Table comparing VCoM estimation accuracy between direct regression from pressure input and SMPL pose-based inference. Metrics are MAE and RMSE. The SMPL pose-based method significantly outperforms the pressure-only approach, achieving lower errors: MAE 31.4 vs. 127.5, and RMSE 42.8 vs. 165.2.}
\caption{VCoM estimation from pressure dynamics: direct regression vs. SMPL mesh-based method. MAE and RMSE show ChairPose's mesh-based approach achieves higher accuracy.}
\begin{tabular}{lcc}
\hline
Input Type & MAE $\downarrow$ & RMSE $\downarrow$ \\
\hline
Pressure & 127.5$\pm$5.3 & 165.2$\pm$6.7 \\
SMPL Pose & \textbf{31.4}$\pm$\textbf{2.1} & \textbf{42.8}$\pm$\textbf{2.8} \\
\hline
\end{tabular}
\label{tab:com_results}
\end{table}

\subsection{Posture Monitoring}
Maintaining proper seated posture is essential for preventing musculoskeletal disorders and promoting long-term health, particularly in the workplace and rehabilitation settings. Fine-grained monitoring of posture can help detect early signs of slouching, forward head posture, or asymmetrical sitting conditions, often linked to discomfort or injury over time. Real-time feedback based on posture analysis can play a crucial role in improving ergonomics and encouraging healthier sitting habits.

ChairPose addresses this need by reconstructing the full-body pose and computing spine angles from the SMPL mesh. This enables estimation of spinal curvature and orientation in both sagittal and coronal planes. Using these measurements, ChairPose can accurately identify deviations from ergonomic postures. Compared to motion capture ground truth, the system achieves a mean absolute angular error of 4.2° for lumbar flexion and 3.8° for thoracic tilt, enabling reliable detection of subtle posture variations. These capabilities make ChairPose suitable for real-time posture assessment tools in occupational health, physical therapy, and smart furniture applications.
\subsection{Improved Action Classification}

\begin{table}[!t]
\Description{Table showing Macro F1 Scores for seated action recognition using ALS-HAR and WS-HAR models across three input types: pressure only, SMPL pose only, and a combination of both. In both models, combining pressure and SMPL pose features yields the highest F1 scores: 0.824 for ALS-HAR and 0.853 for WS-HAR, outperforming single-modality inputs.}
\caption{Macro F1 Scores for seated action recognition across two multi-modal architectures, demonstrating consistent improvement when combining pressure data with SMPL pose features generated by ChairPose, compared to using raw pressure alone.}
\begin{tabular}{lcc}
\hline
Input Type & Model & Macro F1 Score $\uparrow$ \\
\hline
Pressure &  & 0.813$\pm$0.024 \\
SMPL Pose & ALS-HAR\cite{ray2024har} & 0.781$\pm$0.031 \\
SMPL Pose + Pressure &  & \textbf{0.824}$\pm$\textbf{0.022} \\
\hline
Pressure &  & 0.834$\pm$0.017 \\
SMPL Pose & WS-HAR\cite{tarekegn2023enhancing} & 0.819$\pm$0.026 \\
SMPL Pose + Pressure &  & \textbf{0.853}$\pm$\textbf{0.015} \\
\hline
\end{tabular}
\label{tab:classfication_results}
\end{table}

Accurate recognition of seated actions is crucial for a variety of real-world applications, including gesture-based interfaces, assistive technologies, and intelligent workplace systems. However, pressure-only data can be ambiguous, especially for fine-grained or subtle movements. Enhancing this data with body pose information can provide valuable context for distinguishing between similar seated gestures.

To this end, we demonstrate that fusing the full-body pose inferred by ChairPose with the original pressure signals leads to improved action recognition performance. We evaluated this on the TDSD dataset, which includes 12 distinct seated actions, using two sensor fusion models: ALS-HAR \cite{ray2024har} and WS-HAR \cite{tarekegn2023enhancing}. Performance was assessed using the Macro F1 Score, a robust metric for multi-class action recognition. As shown in \cref{tab:classfication_results}, the inclusion of ChairPose-based pose features led to a 12–18\% improvement in F1 Score across models, demonstrating its effectiveness in enabling high-fidelity seated gesture recognition for control and interaction scenarios

\subsection{Example Usecases}

\begin{figure}[!t]
\centering
\includegraphics[width=0.45\textwidth]{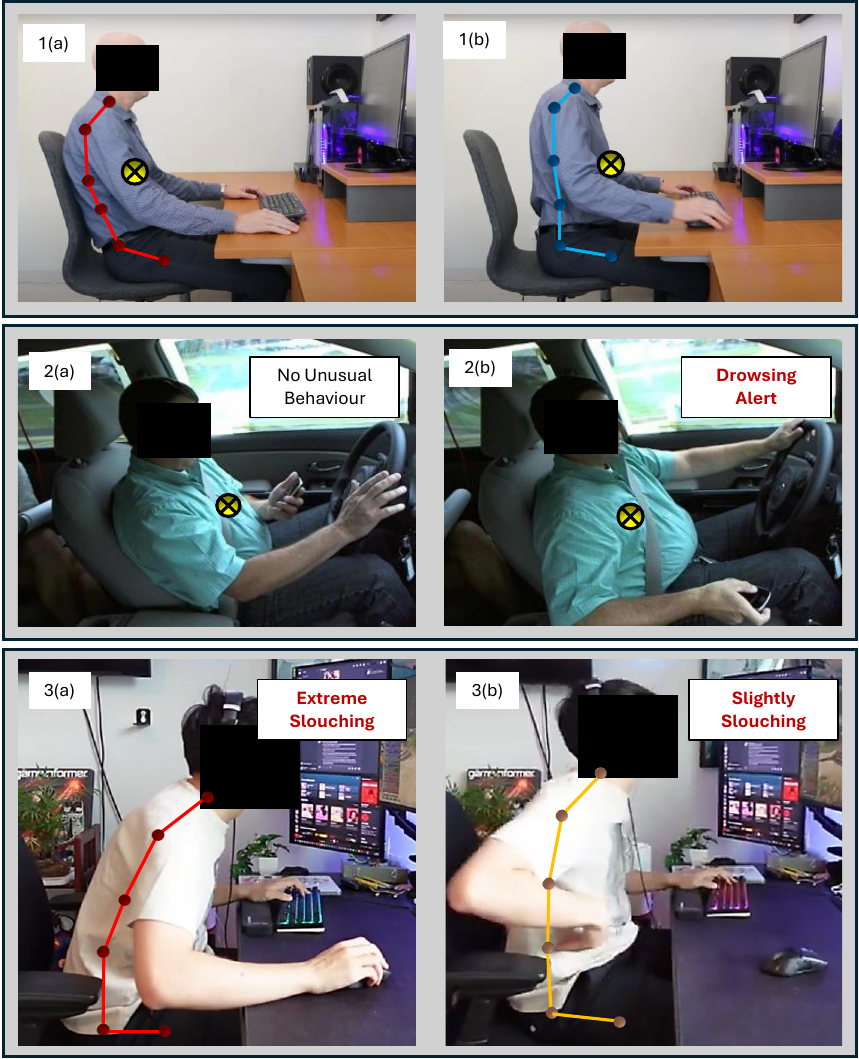} 
\Description{Three example images showing ChairPose applications in real-world settings. Includes posture monitoring in office environments, driver drowsiness detection in vehicles, and gaming posture feedback. Each scenario is annotated with posture assessments such as 'No Unusual Behaviour', 'Drowsing Alert', 'Extreme Slouching', and 'Slightly Slouching', based on spinal alignment and center of mass analysis.}

\caption{Example applications of ChairPose in real-world seated settings.
(1) Workplace Ergonomic Feedback: Provides posture feedback based on spine angle and VCoM tracking to promote healthier sitting habits, reducing the risk of musculoskeletal strain during prolonged desk work.
(2) Driver Monitoring System: Detects Drowsing in real time by continuously estimating the VCoM, enabling early warnings during long drives.
(3) Gaming Posture Alert System: Monitors gamers’ spinal alignment and center of mass to detect sustained bad posture, delivering timely alerts to encourage repositioning and prevent long-term health issues.}
\label{fig:demo}
\end{figure}

By leveraging the hitherto technical benefits, ChairPose can be adopted for a wide range of applications in modern society where people spend a lot of time sitting, for example, at work, dining, commuting, driving, relaxing, etc. We can consider the following example use cases, intended to showcase the broader application space enabled by ChairPose. \cref{fig:demo} illustrates three specific scenarios where ChairPose can enhance comfort, safety, and interactivity in seated environments.

\subsubsection{Sedentary workplace exercises}
ChairPose provides continuous, objective assessment of seated posture for workplace ergonomics and wellness applications. It could enable visual feedback to encourage postural adjustments and breaks, facilitate automated ergonomic evaluations without subjective reporting, and suggest personalized workstation modifications. ChairPose also has the potential to guide users through tailored in-seat exercises. Aggregate data insights may inform broader workplace design improvements aimed at enhancing comfort and reducing fatigue that is informed by the workspace occupants' habits.

\subsubsection{Automotive Transportation} ChairPose presents promising applications for the transportation sector by enabling privacy-preserving monitoring of driver state and occupant safety. Potential uses include detecting drowsiness or inattentiveness via posture analysis, optimizing airbag deployment based on seating position, and dynamically adjusting seats for enhanced comfort. ChairPose could also provide ergonomic insights to promote healthier driving habits, contributing to a safer and more comfortable transportation experience.

\subsubsection{Gaming} ChairPose offers a novel approach to gaming by translating seated posture and center of body mass into intuitive in-game actions, from subtle controls like steering to weight-activated abilities. Beyond enhanced input, it could enable more realistic character embodiment, dynamic adjustment based on player engagement, and the potential for integrating the chair itself into gameplay – creating immersive experiences without relying on cameras or cumbersome motion capture. This overall promotes physically engaging and privacy-respecting gaming.

\subsubsection{Healthcare and Rehabilitation}
While pressure distribution is an established sensing modality for the rehabilitation of wheelchair users, ChairPose offers potential by inferring the posture from pressure distribution, moving beyond simple pressure monitoring and providing a more intuitive understanding of the patients' activities. 
This enables proactive management of pressure ulcers or chronic pain through posture-symptom correlation, tracking of rehabilitation progress, or remote assessment of patients' activity and body positioning.

\section{Discussions \& Limitations}
\subsection{Design Impact}
\revise{The ChairPose sensor mat offers a wearable-free, passive, and unobtrusive design that requires minimal setup—only ensuring the mat’s center is positioned between the seat and backrest—eliminating the need for complex synchronization or calibration. Preliminary user feedback confirmed minimal discomfort or distraction, even during extended sitting periods, with users noting that the mat felt similar to a standard cushion and did not interfere with natural posture or movement. Its sleek, low-profile form factor maintains the chair’s aesthetics without adding visual clutter or bulk.}

\revise{Importantly, ChairPose provides a privacy-preserving alternative to camera-based systems by relying solely on pressure data, which contains no identifiable visual information. This enables deployment in privacy-sensitive environments such as homes, hospitals, and clinics where video monitoring is impractical or unwanted. Unlike wearables, ChairPose avoids any discomfort or compliance issues associated with donning devices, supporting continuous and unobtrusive monitoring.}

\revise{This passive, easy-to-integrate sensor mat enables novel applications beyond the scope of prior embedded sensing systems, including long-term health monitoring, rehabilitation, and ergonomic assessments, all while respecting user comfort and privacy. These features position ChairPose as a scalable, user-friendly solution with significant potential for real-world adoption across diverse settings.}

\subsection{Limitations}

While ChairPose achieves promising results in estimating 3D seated poses from pressure data, several limitations remain that highlight opportunities for further development and future research.

\subsubsection{Deformable Seating Dynamics}
Accurate pose estimation is further challenged by seats made from soft, non-rigid materials that interact dynamically with the human body. These materials can dampen or distort pressure patterns, making it harder for the model to extract meaningful representations.
This issue could benefit from modeling deformable surface dynamics more explicitly—either through enhanced simulation of cushion behavior or by learning latent features that capture deformation-induced variability. Incorporating these dynamics into the training process may improve the model’s robustness to real-world, non-ideal seating conditions.

\subsubsection{Arm and Head Estimation}
While ChairPose provides generally plausible full-body reconstructions, arm and head pose estimation is less accurate in certain cases. This is expected, as they often float above the seat surface and thus contribute minimally to the pressure signal, making it difficult for a purely data-driven model to infer their precise configuration.
This limitation highlights a key constraint of relying solely on pressure-based sensing. One potential solution is to augment ChairPose with additional sensors, such as IMUs, which can provide reliable orientation and motion data for them.

\subsubsection{Error Accumulation}
The P2P model currently employs an autoregressive decoding strategy, which is susceptible to compounding errors over time. Small inaccuracies in early predictions can propagate through the sequence, leading to implausible poses or noticeable drift, especially in scenarios involving movement or pose transitions.
Mitigating this issue could involve alternative decoding strategies or incorporating constraints that enforce physical plausibility over time. For instance, leveraging collision-aware losses based on the known geometry of the chair could help prevent unrealistic body–chair interactions, providing both improved accuracy and physical consistency without sacrificing real-time performance.

\subsection{Responsiveness vs Accuracy}

\begin{figure}[!t]
\centering
\includegraphics[width=0.45\textwidth]{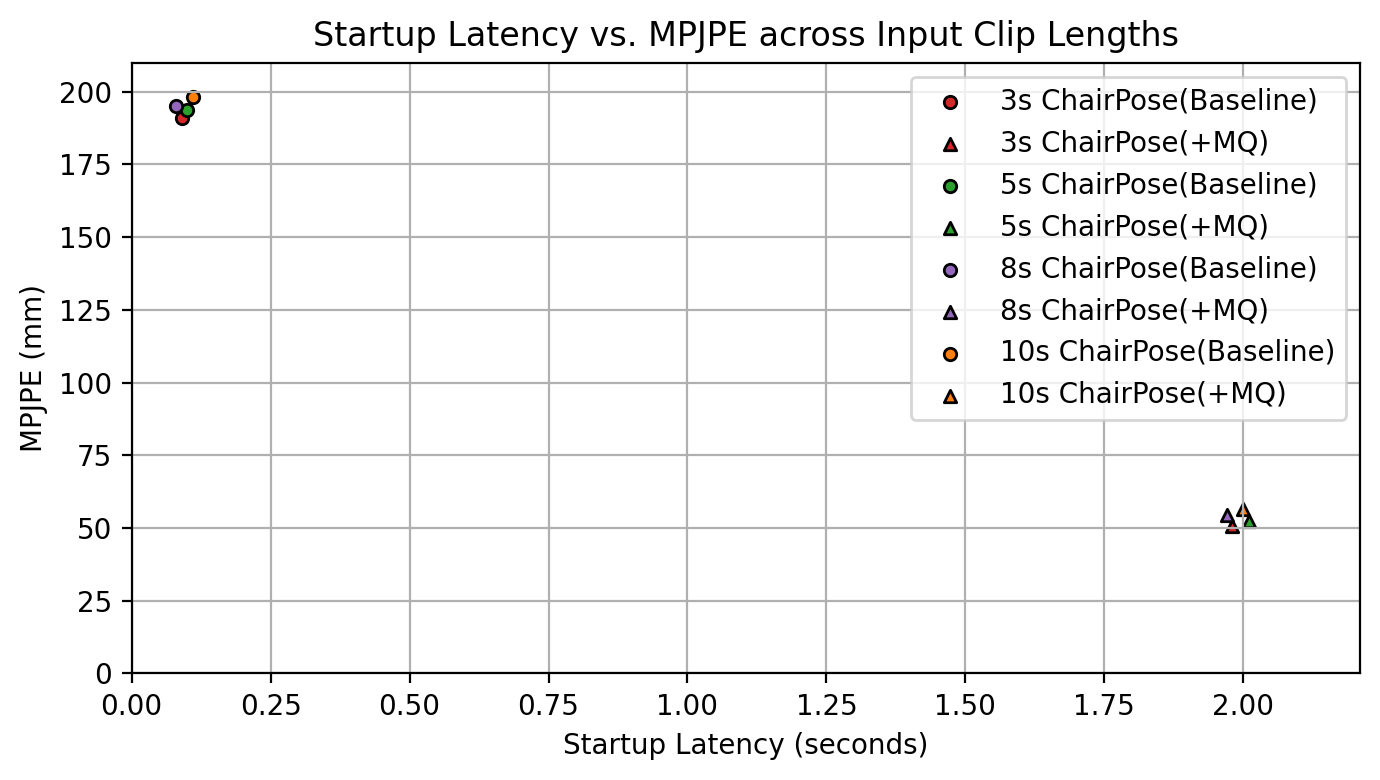} 
\Description{Scatter plot comparing startup latency and mean per joint position error for ChairPose baseline and ChairPose with MQ across various input clip lengths. Baseline methods cluster near low latency but high MPJPE, while MQ-enhanced variants appear at a fixed 2 second latency with significantly lower MPJPE, independent of input length.}
\caption{Latency (startup delay) vs. MPJPE for different input lengths. ChairPose (baseline) offers low latency ($<0.11\,\text{s}$) but high error, while ChairPose (+MQ) achieves much lower MPJPE ($\sim$53\,mm) with a fixed $\sim$2\,sec latency until it generates its first output regardless of input length.} 
\label{fig:lim}
\end{figure}

ChairPose provides two modes of operation reflecting a trade-off between speed and accuracy as shown in \cref{fig:lim}. The fast mode uses the baseline P2P module to directly decode raw motion data, enabling real-time responsiveness with accuracy comparable to state-of-the-art methods. This setup is ideal for interactive applications such as gaming or VR, where low latency is crucial.

ChairPose offers flexible deployment through two operation modes, balancing responsiveness and precision. In time-critical scenarios like interactive gaming or real-time feedback, the fast mode delivers low-latency pose estimates using only the P2P module. For applications demanding higher fidelity, such as clinical assessments or ergonomic evaluations, the full MQ + P2P pipeline offers improved accuracy, at the cost of a fixed 2-second delay due to its temporal receptive field.

The high-accuracy mode combines both the MQ and P2P modules. While this configuration yields more precise pose estimates, it introduces a fixed latency of around two seconds due to the pipeline's temporal receptive field, independent of computational constraints. This mode is better suited for scenarios where accuracy takes precedence, such as healthcare or ergonomic monitoring.

For a well-defined domain-specific application, prior optimization can be made. For example, in a gaming scenario, to insert actions like dodging, leaning, crouching, etc., using only the VcoM could be sufficient. Then the architecture can be modified only to output the 3D coordinate of the VCoM, stripping the full pose estimation parts, thus providing improved accuracy with real-time response.

\section{Conclusion}
In this paper, we introduced ChairPose, a novel, wearable, and vision-free approach for estimating full-body 3D seated human poses from pressure maps, effectively bridging human pose understanding and physical interaction in seated environments. 
By integrating motion quantization and an autoregressive pose generation model, ChairPose enables efficient, temporally coherent, and robust pose estimation without requiring visual sensors, wearable devices, or per-user calibration.

Our evaluation demonstrated that the proposed ChairPose system generalizes effectively across diverse seated behaviors and chair geometries, showing strong performance under both user-specific and novel chair scenarios.
Additionally, our simulation-driven data augmentation pipeline, which leverages only 3D chair scans, significantly reduces the dependency on extensive real-world data collection, yielding accuracy improvements of up to \revise{ 46\%}.

Future work will explore expanding the diversity of users and seating types, and incorporating soft-material dynamics to better capture interactions with deformable surfaces. 
These directions will further strengthen ChairPose's robustness and applicability across ergonomic, assistive, and interactive computing contexts.

\begin{acks}
The research reported in this paper was supported by the BMBF in the project CrossAct (01IW25001) and IITP grant funded by the Korea government(MSIT) (No. RS-2019-II190079).
\end{acks}

\bibliographystyle{ACM-Reference-Format}
\bibliography{sample-base}


\begin{thebibliography}{68}


\ifx \showCODEN    \undefined \def \showCODEN     #1{\unskip}     \fi
\ifx \showISBNx    \undefined \def \showISBNx     #1{\unskip}     \fi
\ifx \showISBNxiii \undefined \def \showISBNxiii  #1{\unskip}     \fi
\ifx \showISSN     \undefined \def \showISSN      #1{\unskip}     \fi
\ifx \showLCCN     \undefined \def \showLCCN      #1{\unskip}     \fi
\ifx \shownote     \undefined \def \shownote      #1{#1}          \fi
\ifx \showarticletitle \undefined \def \showarticletitle #1{#1}   \fi
\ifx \showURL      \undefined \def \showURL       {\relax}        \fi
\providecommand\bibfield[2]{#2}
\providecommand\bibinfo[2]{#2}
\providecommand\natexlab[1]{#1}
\providecommand\showeprint[2][]{arXiv:#2}

\bibitem[Alattas and Elleithy(2014)]%
        {alattas2014detecting}
\bibfield{author}{\bibinfo{person}{Reem Alattas} {and} \bibinfo{person}{Khaled Elleithy}.} \bibinfo{year}{2014}\natexlab{}.
\newblock \showarticletitle{Detecting and minimizing bad posture using postuino among engineering students}. In \bibinfo{booktitle}{\emph{2014 2nd International Conference on Artificial Intelligence, Modelling and Simulation}}. IEEE, \bibinfo{pages}{344--349}.
\newblock


\bibitem[Aminosharieh~Najafi et~al\mbox{.}(2022)]%
        {aminosharieh2022development}
\bibfield{author}{\bibinfo{person}{Taraneh Aminosharieh~Najafi}, \bibinfo{person}{Antonio Abramo}, \bibinfo{person}{Kyandoghere Kyamakya}, {and} \bibinfo{person}{Antonio Affanni}.} \bibinfo{year}{2022}\natexlab{}.
\newblock \showarticletitle{Development of a smart chair sensors system and classification of sitting postures with deep learning algorithms}.
\newblock \bibinfo{journal}{\emph{Sensors}} \bibinfo{volume}{22}, \bibinfo{number}{15} (\bibinfo{year}{2022}), \bibinfo{pages}{5585}.
\newblock


\bibitem[Ardito et~al\mbox{.}(2021)]%
        {ardito2021low}
\bibfield{author}{\bibinfo{person}{Marilda Ardito}, \bibinfo{person}{Fabiana Mascolo}, \bibinfo{person}{Martina Valentini}, {and} \bibinfo{person}{Francesco Dell’Olio}.} \bibinfo{year}{2021}\natexlab{}.
\newblock \showarticletitle{Low-cost wireless wearable system for posture monitoring}.
\newblock \bibinfo{journal}{\emph{Electronics}} \bibinfo{volume}{10}, \bibinfo{number}{21} (\bibinfo{year}{2021}), \bibinfo{pages}{2569}.
\newblock


\bibitem[Ashruf(2002)]%
        {ashruf2002thin}
\bibfield{author}{\bibinfo{person}{CMA Ashruf}.} \bibinfo{year}{2002}\natexlab{}.
\newblock \showarticletitle{Thin flexible pressure sensors}.
\newblock \bibinfo{journal}{\emph{Sensor Review}} \bibinfo{volume}{22}, \bibinfo{number}{4} (\bibinfo{year}{2002}), \bibinfo{pages}{322--327}.
\newblock


\bibitem[Bibbo et~al\mbox{.}(2019)]%
        {bibbo2019sitting}
\bibfield{author}{\bibinfo{person}{Daniele Bibbo}, \bibinfo{person}{Marco Carli}, \bibinfo{person}{Silvia Conforto}, {and} \bibinfo{person}{Federica Battisti}.} \bibinfo{year}{2019}\natexlab{}.
\newblock \showarticletitle{A sitting posture monitoring instrument to assess different levels of cognitive engagement}.
\newblock \bibinfo{journal}{\emph{Sensors}} \bibinfo{volume}{19}, \bibinfo{number}{3} (\bibinfo{year}{2019}), \bibinfo{pages}{455}.
\newblock


\bibitem[Cai et~al\mbox{.}(2023)]%
        {cai2023smpler}
\bibfield{author}{\bibinfo{person}{Zhongang Cai}, \bibinfo{person}{Wanqi Yin}, \bibinfo{person}{Ailing Zeng}, \bibinfo{person}{Chen Wei}, \bibinfo{person}{Qingping Sun}, \bibinfo{person}{Wang Yanjun}, \bibinfo{person}{Hui~En Pang}, \bibinfo{person}{Haiyi Mei}, \bibinfo{person}{Mingyuan Zhang}, \bibinfo{person}{Lei Zhang}, {et~al\mbox{.}}} \bibinfo{year}{2023}\natexlab{}.
\newblock \showarticletitle{Smpler-x: Scaling up expressive human pose and shape estimation}.
\newblock \bibinfo{journal}{\emph{Advances in Neural Information Processing Systems}}  \bibinfo{volume}{36} (\bibinfo{year}{2023}), \bibinfo{pages}{11454--11468}.
\newblock


\bibitem[Celik et~al\mbox{.}(2018)]%
        {celik2018determination}
\bibfield{author}{\bibinfo{person}{Sevim Celik}, \bibinfo{person}{Kadir Celik}, \bibinfo{person}{EL{\.I}F D{\.I}R{\.I}ME{\c{S}}E}, \bibinfo{person}{Nurten Ta{\c{s}}demir}, \bibinfo{person}{Tarik Arik}, {and} \bibinfo{person}{Ibrahim Buyukkara}.} \bibinfo{year}{2018}\natexlab{}.
\newblock \showarticletitle{Determination of pain in musculoskeletal system reported by office workers and the pain risk factors}.
\newblock \bibinfo{journal}{\emph{International journal of occupational medicine and environmental health}} \bibinfo{volume}{31}, \bibinfo{number}{1} (\bibinfo{year}{2018}).
\newblock


\bibitem[Chen et~al\mbox{.}(2025)]%
        {chen2025skeleton}
\bibfield{author}{\bibinfo{person}{Kaixin Chen}, \bibinfo{person}{Lin Zhang}, \bibinfo{person}{Zhong Wang}, \bibinfo{person}{Shengjie Zhao}, {and} \bibinfo{person}{Yicong Zhou}.} \bibinfo{year}{2025}\natexlab{}.
\newblock \showarticletitle{Skeleton-aware graph-based adversarial networks for human pose estimation from sparse IMUs}.
\newblock \bibinfo{journal}{\emph{ACM Transactions on Multimedia Computing, Communications and Applications}} \bibinfo{volume}{21}, \bibinfo{number}{4} (\bibinfo{year}{2025}), \bibinfo{pages}{1--22}.
\newblock


\bibitem[Chen et~al\mbox{.}(2024a)]%
        {chen2024cavatar}
\bibfield{author}{\bibinfo{person}{Wenqiang Chen}, \bibinfo{person}{Yexin Hu}, \bibinfo{person}{Wei Song}, \bibinfo{person}{Yingcheng Liu}, \bibinfo{person}{Antonio Torralba}, {and} \bibinfo{person}{Wojciech Matusik}.} \bibinfo{year}{2024}\natexlab{a}.
\newblock \showarticletitle{CAvatar: Real-time Human Activity Mesh Reconstruction via Tactile Carpets}.
\newblock \bibinfo{journal}{\emph{Proceedings of the ACM on Interactive, Mobile, Wearable and Ubiquitous Technologies}} \bibinfo{volume}{7}, \bibinfo{number}{4} (\bibinfo{year}{2024}), \bibinfo{pages}{1--24}.
\newblock


\bibitem[Chen et~al\mbox{.}(2024b)]%
        {chen2024sato}
\bibfield{author}{\bibinfo{person}{Wenshuo Chen}, \bibinfo{person}{Hongru Xiao}, \bibinfo{person}{Erhang Zhang}, \bibinfo{person}{Lijie Hu}, \bibinfo{person}{Lei Wang}, \bibinfo{person}{Mengyuan Liu}, {and} \bibinfo{person}{Chen Chen}.} \bibinfo{year}{2024}\natexlab{b}.
\newblock \showarticletitle{SATO: Stable Text-to-Motion Framework}. In \bibinfo{booktitle}{\emph{Proceedings of the 32nd ACM International Conference on Multimedia}}. \bibinfo{pages}{6989--6997}.
\newblock


\bibitem[Clever et~al\mbox{.}(2022)]%
        {clever2022bodypressure}
\bibfield{author}{\bibinfo{person}{Henry~M Clever}, \bibinfo{person}{Patrick~L Grady}, \bibinfo{person}{Greg Turk}, {and} \bibinfo{person}{Charles~C Kemp}.} \bibinfo{year}{2022}\natexlab{}.
\newblock \showarticletitle{Bodypressure-inferring body pose and contact pressure from a depth image}.
\newblock \bibinfo{journal}{\emph{IEEE Transactions on Pattern Analysis and Machine Intelligence}} \bibinfo{volume}{45}, \bibinfo{number}{1} (\bibinfo{year}{2022}), \bibinfo{pages}{137--153}.
\newblock


\bibitem[Daneshmandi et~al\mbox{.}(2017)]%
        {daneshmandi2017adverse}
\bibfield{author}{\bibinfo{person}{Hadi Daneshmandi}, \bibinfo{person}{Alireza Choobineh}, \bibinfo{person}{Haleh Ghaem}, {and} \bibinfo{person}{Mehran Karimi}.} \bibinfo{year}{2017}\natexlab{}.
\newblock \showarticletitle{Adverse effects of prolonged sitting behavior on the general health of office workers}.
\newblock \bibinfo{journal}{\emph{Journal of lifestyle medicine}} \bibinfo{volume}{7}, \bibinfo{number}{2} (\bibinfo{year}{2017}), \bibinfo{pages}{69}.
\newblock


\bibitem[Demmans et~al\mbox{.}(2007)]%
        {demmans2007posture}
\bibfield{author}{\bibinfo{person}{Carrie Demmans}, \bibinfo{person}{Sriram Subramanian}, {and} \bibinfo{person}{Jon Titus}.} \bibinfo{year}{2007}\natexlab{}.
\newblock \showarticletitle{Posture monitoring and improvement for laptop use}. In \bibinfo{booktitle}{\emph{CHI'07 Extended Abstracts on Human Factors in Computing Systems}}. \bibinfo{pages}{2357--2362}.
\newblock


\bibitem[Fan et~al\mbox{.}(2024)]%
        {fan2024freemotion}
\bibfield{author}{\bibinfo{person}{Ke Fan}, \bibinfo{person}{Junshu Tang}, \bibinfo{person}{Weijian Cao}, \bibinfo{person}{Ran Yi}, \bibinfo{person}{Moran Li}, \bibinfo{person}{Jingyu Gong}, \bibinfo{person}{Jiangning Zhang}, \bibinfo{person}{Yabiao Wang}, \bibinfo{person}{Chengjie Wang}, {and} \bibinfo{person}{Lizhuang Ma}.} \bibinfo{year}{2024}\natexlab{}.
\newblock \showarticletitle{Freemotion: A unified framework for number-free text-to-motion synthesis}. In \bibinfo{booktitle}{\emph{European Conference on Computer Vision}}. Springer, \bibinfo{pages}{93--109}.
\newblock


\bibitem[Feng et~al\mbox{.}(2019)]%
        {feng2019you}
\bibfield{author}{\bibinfo{person}{Lin Feng}, \bibinfo{person}{Ziyi Li}, {and} \bibinfo{person}{Chen Liu}.} \bibinfo{year}{2019}\natexlab{}.
\newblock \showarticletitle{Are you sitting right?-sitting posture recognition using RF signals}. In \bibinfo{booktitle}{\emph{2019 IEEE Pacific Rim Conference on Communications, Computers and Signal Processing (PACRIM)}}. IEEE, \bibinfo{pages}{1--6}.
\newblock


\bibitem[Feng et~al\mbox{.}(2020)]%
        {feng2020sitr}
\bibfield{author}{\bibinfo{person}{Lin Feng}, \bibinfo{person}{Ziyi Li}, \bibinfo{person}{Chen Liu}, \bibinfo{person}{Xiaojiang Chen}, \bibinfo{person}{Xiao Yin}, {and} \bibinfo{person}{Dingyi Fang}.} \bibinfo{year}{2020}\natexlab{}.
\newblock \showarticletitle{SitR: Sitting posture recognition using RF signals}.
\newblock \bibinfo{journal}{\emph{IEEE Internet of Things Journal}} \bibinfo{volume}{7}, \bibinfo{number}{12} (\bibinfo{year}{2020}), \bibinfo{pages}{11492--11504}.
\newblock


\bibitem[Goswami et~al\mbox{.}(2024)]%
        {goswami2024hypervq}
\bibfield{author}{\bibinfo{person}{Nabarun Goswami}, \bibinfo{person}{Yusuke Mukuta}, {and} \bibinfo{person}{Tatsuya Harada}.} \bibinfo{year}{2024}\natexlab{}.
\newblock \showarticletitle{Hypervq: Mlr-based vector quantization in hyperbolic space}.
\newblock \bibinfo{journal}{\emph{arXiv preprint arXiv:2403.13015}} (\bibinfo{year}{2024}).
\newblock


\bibitem[Greenwood-Hickman et~al\mbox{.}(2021)]%
        {greenwood2021cnn}
\bibfield{author}{\bibinfo{person}{Mikael~Anne Greenwood-Hickman}, \bibinfo{person}{Supun Nakandala}, \bibinfo{person}{Marta~M Jankowska}, \bibinfo{person}{Dori~E Rosenberg}, \bibinfo{person}{Fatima Tuz-Zahra}, \bibinfo{person}{John Bellettiere}, \bibinfo{person}{Jordan Carlson}, \bibinfo{person}{Paul~R Hibbing}, \bibinfo{person}{Jingjing Zou}, \bibinfo{person}{Andrea~Z Lacroix}, {et~al\mbox{.}}} \bibinfo{year}{2021}\natexlab{}.
\newblock \showarticletitle{The CNN Hip Accelerometer Posture (CHAP) method for classifying sitting patterns from hip accelerometers: A validation study}.
\newblock \bibinfo{journal}{\emph{Medicine and science in sports and exercise}} \bibinfo{volume}{53}, \bibinfo{number}{11} (\bibinfo{year}{2021}), \bibinfo{pages}{2445}.
\newblock


\bibitem[Guo et~al\mbox{.}(2022)]%
        {guo2022generating}
\bibfield{author}{\bibinfo{person}{Chuan Guo}, \bibinfo{person}{Shihao Zou}, \bibinfo{person}{Xinxin Zuo}, \bibinfo{person}{Sen Wang}, \bibinfo{person}{Wei Ji}, \bibinfo{person}{Xingyu Li}, {and} \bibinfo{person}{Li Cheng}.} \bibinfo{year}{2022}\natexlab{}.
\newblock \showarticletitle{Generating diverse and natural 3d human motions from text}. In \bibinfo{booktitle}{\emph{Proceedings of the IEEE/CVF conference on computer vision and pattern recognition}}. \bibinfo{pages}{5152--5161}.
\newblock


\bibitem[Gwak et~al\mbox{.}(2024)]%
        {gwak2024physiological}
\bibfield{author}{\bibinfo{person}{Jongseong Gwak}, \bibinfo{person}{Kazuyoshi Arata}, \bibinfo{person}{Takumi Yamakawa}, \bibinfo{person}{Hideo Tobata}, \bibinfo{person}{Motoki Shino}, {and} \bibinfo{person}{Yoshihiro Suda}.} \bibinfo{year}{2024}\natexlab{}.
\newblock \showarticletitle{Physiological Responses Related to Sitting Comfort Due to Changes in Seat Parameters}.
\newblock \bibinfo{journal}{\emph{Applied Sciences}} \bibinfo{volume}{14}, \bibinfo{number}{17} (\bibinfo{year}{2024}), \bibinfo{pages}{7870}.
\newblock


\bibitem[Han et~al\mbox{.}({[n.\,d.]})]%
        {hansmart}
\bibfield{author}{\bibinfo{person}{Isaac Han}, \bibinfo{person}{Seoyoung Lee}, \bibinfo{person}{Sangyeon Park}, \bibinfo{person}{Ecehan Akan}, \bibinfo{person}{Yiyue Luo}, {and} \bibinfo{person}{Kyung-Joong Kim}.} \bibinfo{year}{[n.\,d.]}\natexlab{}.
\newblock \showarticletitle{Smart Insole: Predicting 3D human pose from foot pressure}. In \bibinfo{booktitle}{\emph{2nd NeurIPS Workshop on Touch Processing: From Data to Knowledge}}.
\newblock


\bibitem[Huang and Zhou(2024)]%
        {huang2024scalable}
\bibfield{author}{\bibinfo{person}{Jiawei Huang} {and} \bibinfo{person}{Ding Zhou}.} \bibinfo{year}{2024}\natexlab{}.
\newblock \showarticletitle{A scalable real-time computer vision system for student posture detection in smart classrooms}.
\newblock \bibinfo{journal}{\emph{Education and Information Technologies}} \bibinfo{volume}{29}, \bibinfo{number}{1} (\bibinfo{year}{2024}), \bibinfo{pages}{917--937}.
\newblock


\bibitem[Jin et~al\mbox{.}(2025)]%
        {jin2025sitpose}
\bibfield{author}{\bibinfo{person}{Hang Jin}, \bibinfo{person}{Xin He}, \bibinfo{person}{Lingyun Wang}, \bibinfo{person}{Yujun Zhu}, \bibinfo{person}{Weiwei Jiang}, {and} \bibinfo{person}{Xiaobo Zhou}.} \bibinfo{year}{2025}\natexlab{}.
\newblock \showarticletitle{SitPose: Real-Time Detection of Sitting Posture and Sedentary Behavior Using Ensemble Learning With Depth Sensor}.
\newblock \bibinfo{journal}{\emph{IEEE Sensors Journal}} (\bibinfo{year}{2025}).
\newblock


\bibitem[Krauter et~al\mbox{.}(2024)]%
        {krauter2024sitting}
\bibfield{author}{\bibinfo{person}{Christian Krauter}, \bibinfo{person}{Katrin Angerbauer}, \bibinfo{person}{Aim{\'e}e Sousa~Calepso}, \bibinfo{person}{Alexander Achberger}, \bibinfo{person}{Sven Mayer}, {and} \bibinfo{person}{Michael Sedlmair}.} \bibinfo{year}{2024}\natexlab{}.
\newblock \showarticletitle{Sitting posture recognition and feedback: a literature review}. In \bibinfo{booktitle}{\emph{Proceedings of the 2024 CHI Conference on Human Factors in Computing Systems}}. \bibinfo{pages}{1--20}.
\newblock


\bibitem[Li et~al\mbox{.}(2023)]%
        {li2023abnormal}
\bibfield{author}{\bibinfo{person}{Linhan Li}, \bibinfo{person}{Guanci Yang}, \bibinfo{person}{Yang Li}, \bibinfo{person}{Dongying Zhu}, {and} \bibinfo{person}{Ling He}.} \bibinfo{year}{2023}\natexlab{}.
\newblock \showarticletitle{Abnormal sitting posture recognition based on multi-scale spatiotemporal features of skeleton graph}.
\newblock \bibinfo{journal}{\emph{Engineering Applications of Artificial Intelligence}}  \bibinfo{volume}{123} (\bibinfo{year}{2023}), \bibinfo{pages}{106374}.
\newblock


\bibitem[Li et~al\mbox{.}(2024a)]%
        {li2024semantically}
\bibfield{author}{\bibinfo{person}{Weijian Li}, \bibinfo{person}{Shijie Li}, {and} \bibinfo{person}{Huaiguang Jiang}.} \bibinfo{year}{2024}\natexlab{a}.
\newblock \showarticletitle{Semantically Encoding Enhancements for Unsupervised Color Feature Separation}. In \bibinfo{booktitle}{\emph{2024 IEEE International Conference on Smart City (SmartCity)}}. IEEE, \bibinfo{pages}{77--82}.
\newblock


\bibitem[Li et~al\mbox{.}(2024b)]%
        {li2024llama}
\bibfield{author}{\bibinfo{person}{Yanwei Li}, \bibinfo{person}{Chengyao Wang}, {and} \bibinfo{person}{Jiaya Jia}.} \bibinfo{year}{2024}\natexlab{b}.
\newblock \showarticletitle{Llama-vid: An image is worth 2 tokens in large language models}. In \bibinfo{booktitle}{\emph{European Conference on Computer Vision}}. Springer, \bibinfo{pages}{323--340}.
\newblock


\bibitem[Loper et~al\mbox{.}(2023)]%
        {loper2023smpl}
\bibfield{author}{\bibinfo{person}{Matthew Loper}, \bibinfo{person}{Naureen Mahmood}, \bibinfo{person}{Javier Romero}, \bibinfo{person}{Gerard Pons-Moll}, {and} \bibinfo{person}{Michael~J Black}.} \bibinfo{year}{2023}\natexlab{}.
\newblock \showarticletitle{SMPL: A skinned multi-person linear model}.
\newblock In \bibinfo{booktitle}{\emph{Seminal Graphics Papers: Pushing the Boundaries, Volume 2}}. \bibinfo{pages}{851--866}.
\newblock


\bibitem[Luo et~al\mbox{.}(2021)]%
        {9577856}
\bibfield{author}{\bibinfo{person}{Yiyue Luo}, \bibinfo{person}{Yunzhu Li}, \bibinfo{person}{Michael Foshey}, \bibinfo{person}{Wan Shou}, \bibinfo{person}{Pratyusha Sharma}, \bibinfo{person}{Tomás Palacios}, \bibinfo{person}{Antonio Torralba}, {and} \bibinfo{person}{Wojciech Matusik}.} \bibinfo{year}{2021}\natexlab{}.
\newblock \showarticletitle{Intelligent Carpet: Inferring 3D Human Pose from Tactile Signals}. In \bibinfo{booktitle}{\emph{2021 IEEE/CVF Conference on Computer Vision and Pattern Recognition (CVPR)}}. \bibinfo{pages}{11250--11260}.
\newblock
\href{https://doi.org/10.1109/CVPR46437.2021.01110}{doi:\nolinkurl{10.1109/CVPR46437.2021.01110}}


\bibitem[Mclaughlin et~al\mbox{.}(2020)]%
        {mclaughlin2020worldwide}
\bibfield{author}{\bibinfo{person}{Matthew Mclaughlin}, \bibinfo{person}{AJ Atkin}, \bibinfo{person}{L Starr}, \bibinfo{person}{A Hall}, \bibinfo{person}{L Wolfenden}, \bibinfo{person}{R Sutherland}, \bibinfo{person}{J Wiggers}, \bibinfo{person}{A Ramirez}, \bibinfo{person}{P Hallal}, \bibinfo{person}{M Pratt}, {et~al\mbox{.}}} \bibinfo{year}{2020}\natexlab{}.
\newblock \showarticletitle{Worldwide surveillance of self-reported sitting time: a scoping review}.
\newblock \bibinfo{journal}{\emph{International Journal of Behavioral Nutrition and Physical Activity}} \bibinfo{volume}{17}, \bibinfo{number}{1} (\bibinfo{year}{2020}), \bibinfo{pages}{111}.
\newblock


\bibitem[Mizumoto et~al\mbox{.}(2020)]%
        {mizumoto2020design}
\bibfield{author}{\bibinfo{person}{Teruhiro Mizumoto}, \bibinfo{person}{Yasuhiro Otoda}, \bibinfo{person}{Chihiro Nakajima}, \bibinfo{person}{Mitsuhiro Kohana}, \bibinfo{person}{Motohiro Uenishi}, \bibinfo{person}{Keiichi Yasumoto}, {and} \bibinfo{person}{Yutaka Arakawa}.} \bibinfo{year}{2020}\natexlab{}.
\newblock \showarticletitle{Design and implementation of sensor-embedded chair for continuous sitting posture recognition}.
\newblock \bibinfo{journal}{\emph{IEICE TRANSACTIONS on Information and Systems}} \bibinfo{volume}{103}, \bibinfo{number}{5} (\bibinfo{year}{2020}), \bibinfo{pages}{1067--1077}.
\newblock


\bibitem[Mollyn et~al\mbox{.}(2023)]%
        {mollyn2023imuposer}
\bibfield{author}{\bibinfo{person}{Vimal Mollyn}, \bibinfo{person}{Riku Arakawa}, \bibinfo{person}{Mayank Goel}, \bibinfo{person}{Chris Harrison}, {and} \bibinfo{person}{Karan Ahuja}.} \bibinfo{year}{2023}\natexlab{}.
\newblock \showarticletitle{Imuposer: Full-body pose estimation using imus in phones, watches, and earbuds}. In \bibinfo{booktitle}{\emph{Proceedings of the 2023 CHI Conference on Human Factors in Computing Systems}}. \bibinfo{pages}{1--12}.
\newblock


\bibitem[Mutlu et~al\mbox{.}(2007)]%
        {mutlu2007robust}
\bibfield{author}{\bibinfo{person}{Bilge Mutlu}, \bibinfo{person}{Andreas Krause}, \bibinfo{person}{Jodi Forlizzi}, \bibinfo{person}{Carlos Guestrin}, {and} \bibinfo{person}{Jessica Hodgins}.} \bibinfo{year}{2007}\natexlab{}.
\newblock \showarticletitle{Robust, low-cost, non-intrusive sensing and recognition of seated postures}. In \bibinfo{booktitle}{\emph{Proceedings of the 20th annual ACM symposium on User interface software and technology}}. \bibinfo{pages}{149--158}.
\newblock


\bibitem[Nadeem et~al\mbox{.}(2024)]%
        {nadeem2024sitting}
\bibfield{author}{\bibinfo{person}{Muhammad Nadeem}, \bibinfo{person}{Ersin Elbasi}, \bibinfo{person}{Aymen~I Zreikat}, {and} \bibinfo{person}{Mohammad Sharsheer}.} \bibinfo{year}{2024}\natexlab{}.
\newblock \showarticletitle{Sitting posture recognition systems: Comprehensive literature review and analysis}.
\newblock \bibinfo{journal}{\emph{Applied Sciences}} \bibinfo{volume}{14}, \bibinfo{number}{18} (\bibinfo{year}{2024}), \bibinfo{pages}{8557}.
\newblock


\bibitem[Nishimura et~al\mbox{.}(2023)]%
        {nishimura2023detection}
\bibfield{author}{\bibinfo{person}{Kento Nishimura}, \bibinfo{person}{Kodai Ito}, \bibinfo{person}{Ken Fujiwara}, \bibinfo{person}{Kazuyuki Fujita}, {and} \bibinfo{person}{Yuichi Itoh}.} \bibinfo{year}{2023}\natexlab{}.
\newblock \showarticletitle{Detection of nodding of interlocutors using a chair-shaped device and investigating relationship between a divergent thinking task and amount of nodding}.
\newblock \bibinfo{journal}{\emph{Quality and User Experience}} \bibinfo{volume}{8}, \bibinfo{number}{1} (\bibinfo{year}{2023}), \bibinfo{pages}{10}.
\newblock


\bibitem[Nouriani et~al\mbox{.}(2024)]%
        {nouriani2024vector}
\bibfield{author}{\bibinfo{person}{Ali Nouriani}, \bibinfo{person}{RA McGovern}, {and} \bibinfo{person}{R Rajamani}.} \bibinfo{year}{2024}\natexlab{}.
\newblock \showarticletitle{Vector-based Inertial Poser: Human pose estimation with high gain observer and deep learning using sparse IMU sensors}.
\newblock \bibinfo{journal}{\emph{Biomedical Signal Processing and Control}}  \bibinfo{volume}{95} (\bibinfo{year}{2024}), \bibinfo{pages}{106432}.
\newblock


\bibitem[Ottosson(2024)]%
        {RagdollDynamics}
\bibfield{author}{\bibinfo{person}{Marcus Ottosson}.} \bibinfo{year}{2024}\natexlab{}.
\newblock \bibinfo{title}{Ragdoll Dynamics}.
\newblock \bibinfo{howpublished}{\url{https://ragdolldynamics.com/}}.
\newblock
\newblock
\shownote{Accessed: 2025-04-09}.


\bibitem[Ramani et~al\mbox{.}(2024)]%
        {ramani2024imuoptimize}
\bibfield{author}{\bibinfo{person}{Varun Ramani}, \bibinfo{person}{Hossein Khayami}, \bibinfo{person}{Yang Bai}, \bibinfo{person}{Nakul Garg}, {and} \bibinfo{person}{Nirupam Roy}.} \bibinfo{year}{2024}\natexlab{}.
\newblock \showarticletitle{IMUOptimize: A Data-Driven Approach to Optimal IMU Placement for Human Pose Estimation with Transformer Architecture}.
\newblock \bibinfo{journal}{\emph{arXiv preprint arXiv:2402.08923}} (\bibinfo{year}{2024}).
\newblock


\bibitem[Ray et~al\mbox{.}(2024a)]%
        {ray2024har}
\bibfield{author}{\bibinfo{person}{Lala Shakti~Swarup Ray}, \bibinfo{person}{Daniel Gei{\ss}ler}, \bibinfo{person}{Mengxi Liu}, \bibinfo{person}{Bo Zhou}, \bibinfo{person}{Sungho Suh}, {and} \bibinfo{person}{Paul Lukowicz}.} \bibinfo{year}{2024}\natexlab{a}.
\newblock \showarticletitle{ALS-HAR: Harnessing Wearable Ambient Light Sensors to Enhance IMU-Based Human Activity Recognition}. In \bibinfo{booktitle}{\emph{International Conference on Pattern Recognition}}. Springer, \bibinfo{pages}{133--147}.
\newblock


\bibitem[Ray et~al\mbox{.}(2025a)]%
        {ray2025txp}
\bibfield{author}{\bibinfo{person}{Lala Shakti~Swarup Ray}, \bibinfo{person}{Lars Krupp}, \bibinfo{person}{Vitor~Fortes Rey}, \bibinfo{person}{Bo Zhou}, \bibinfo{person}{Sungho Suh}, {and} \bibinfo{person}{Paul Lukowicz}.} \bibinfo{year}{2025}\natexlab{a}.
\newblock \showarticletitle{TxP: Reciprocal Generation of Ground Pressure Dynamics and Activity Descriptions for Improving Human Activity Recognition}.
\newblock \bibinfo{journal}{\emph{Proceedings of the ACM on Interactive, Mobile, Wearable and Ubiquitous Technologies}} \bibinfo{volume}{9}, \bibinfo{number}{2} (\bibinfo{year}{2025}), \bibinfo{pages}{1--32}.
\newblock


\bibitem[Ray et~al\mbox{.}(2024c)]%
        {ray2024text}
\bibfield{author}{\bibinfo{person}{Lala Shakti~Swarup Ray}, \bibinfo{person}{Bo Zhou}, \bibinfo{person}{Sungho Suh}, \bibinfo{person}{Lars Krupp}, \bibinfo{person}{Vitor~Fortes Rey}, {and} \bibinfo{person}{Paul Lukowicz}.} \bibinfo{year}{2024}\natexlab{c}.
\newblock \showarticletitle{Text me the data: Generating ground pressure sequence from textual descriptions for har}. In \bibinfo{booktitle}{\emph{2024 IEEE International Conference on Pervasive Computing and Communications Workshops and other Affiliated Events (PerCom Workshops)}}. IEEE, \bibinfo{pages}{461--464}.
\newblock


\bibitem[Ray et~al\mbox{.}(2023a)]%
        {ray2023pressim}
\bibfield{author}{\bibinfo{person}{Lala Shakti~Swarup Ray}, \bibinfo{person}{Bo Zhou}, \bibinfo{person}{Sungho Suh}, {and} \bibinfo{person}{Paul Lukowicz}.} \bibinfo{year}{2023}\natexlab{a}.
\newblock \showarticletitle{Pressim: An end-to-end framework for dynamic ground pressure profile generation from monocular videos using physics-based 3d simulation}. In \bibinfo{booktitle}{\emph{2023 IEEE International Conference on Pervasive Computing and Communications Workshops and other Affiliated Events (PerCom Workshops)}}. IEEE, \bibinfo{pages}{484--489}.
\newblock


\bibitem[Ray et~al\mbox{.}(2023b)]%
        {ray2023selecting}
\bibfield{author}{\bibinfo{person}{Lala Shakti~Swarup Ray}, \bibinfo{person}{Bo Zhou}, \bibinfo{person}{Sungho Suh}, {and} \bibinfo{person}{Paul Lukowicz}.} \bibinfo{year}{2023}\natexlab{b}.
\newblock \showarticletitle{Selecting the motion ground truth for loose-fitting wearables: Benchmarking optical mocap methods}. In \bibinfo{booktitle}{\emph{Proceedings of the 2023 ACM International Symposium on Wearable Computers}}. \bibinfo{pages}{27--32}.
\newblock


\bibitem[Ray et~al\mbox{.}(2024b)]%
        {ray2024comprehensive}
\bibfield{author}{\bibinfo{person}{Lala Shakti~Swarup Ray}, \bibinfo{person}{Bo Zhou}, \bibinfo{person}{Sungho Suh}, {and} \bibinfo{person}{Paul Lukowicz}.} \bibinfo{year}{2024}\natexlab{b}.
\newblock \showarticletitle{A comprehensive evaluation of marker-based, markerless methods for loose garment scenarios in varying camera configurations}.
\newblock \bibinfo{journal}{\emph{Frontiers in Computer Science}}  \bibinfo{volume}{6} (\bibinfo{year}{2024}), \bibinfo{pages}{1379925}.
\newblock


\bibitem[Ray et~al\mbox{.}(2025b)]%
        {10890689}
\bibfield{author}{\bibinfo{person}{Lala Shakti~Swarup Ray}, \bibinfo{person}{Bo Zhou}, \bibinfo{person}{Sungho Suh}, {and} \bibinfo{person}{Paul Lukowicz}.} \bibinfo{year}{2025}\natexlab{b}.
\newblock \showarticletitle{OV-HHIR: Open Vocabulary Human Interaction Recognition Using Cross-modal Integration of Large Language Models}. In \bibinfo{booktitle}{\emph{ICASSP 2025 - 2025 IEEE International Conference on Acoustics, Speech and Signal Processing (ICASSP)}}. \bibinfo{pages}{1--5}.
\newblock
\href{https://doi.org/10.1109/ICASSP49660.2025.10890689}{doi:\nolinkurl{10.1109/ICASSP49660.2025.10890689}}


\bibitem[Roynarin et~al\mbox{.}(2024)]%
        {roynarin2024postural}
\bibfield{author}{\bibinfo{person}{Narumon Roynarin}, \bibinfo{person}{Sirinant Channak}, {and} \bibinfo{person}{Prawit Janwantanakul}.} \bibinfo{year}{2024}\natexlab{}.
\newblock \showarticletitle{Postural shifts and body perceived discomfort during 1-hour sitting when leaning and sitting on an air-filled seat cushion among healthy office workers}.
\newblock \bibinfo{journal}{\emph{Ergonomics}} \bibinfo{volume}{67}, \bibinfo{number}{12} (\bibinfo{year}{2024}), \bibinfo{pages}{2126--2137}.
\newblock


\bibitem[Scott et~al\mbox{.}(2020)]%
        {scott2020image}
\bibfield{author}{\bibinfo{person}{Jesse Scott}, \bibinfo{person}{Bharadwaj Ravichandran}, \bibinfo{person}{Christopher Funk}, \bibinfo{person}{Robert~T Collins}, {and} \bibinfo{person}{Yanxi Liu}.} \bibinfo{year}{2020}\natexlab{}.
\newblock \showarticletitle{From image to stability: Learning dynamics from human pose}. In \bibinfo{booktitle}{\emph{Computer Vision--ECCV 2020: 16th European Conference, Glasgow, UK, August 23--28, 2020, Proceedings, Part XXIII 16}}. Springer, \bibinfo{pages}{536--554}.
\newblock


\bibitem[Seong et~al\mbox{.}(2024)]%
        {seong2024intelligent}
\bibfield{author}{\bibinfo{person}{Minwoo Seong}, \bibinfo{person}{Gwangbin Kim}, \bibinfo{person}{Jaehee Lee}, \bibinfo{person}{Joseph DelPreto}, \bibinfo{person}{Wojciech Matusik}, \bibinfo{person}{Daniela Rus}, {and} \bibinfo{person}{SeungJun Kim}.} \bibinfo{year}{2024}\natexlab{}.
\newblock \showarticletitle{Intelligent Seat: Tactile Signal-Based 3D Sitting Pose Inference}. In \bibinfo{booktitle}{\emph{Companion of the 2024 on ACM International Joint Conference on Pervasive and Ubiquitous Computing}}. \bibinfo{pages}{791--796}.
\newblock


\bibitem[Shimada et~al\mbox{.}(2020)]%
        {shimada2020physcap}
\bibfield{author}{\bibinfo{person}{Soshi Shimada}, \bibinfo{person}{Vladislav Golyanik}, \bibinfo{person}{Weipeng Xu}, {and} \bibinfo{person}{Christian Theobalt}.} \bibinfo{year}{2020}\natexlab{}.
\newblock \showarticletitle{Physcap: Physically plausible monocular 3d motion capture in real time}.
\newblock \bibinfo{journal}{\emph{ACM Transactions on Graphics (ToG)}} \bibinfo{volume}{39}, \bibinfo{number}{6} (\bibinfo{year}{2020}), \bibinfo{pages}{1--16}.
\newblock


\bibitem[Shin et~al\mbox{.}(2024)]%
        {shin2024wham}
\bibfield{author}{\bibinfo{person}{Soyong Shin}, \bibinfo{person}{Juyong Kim}, \bibinfo{person}{Eni Halilaj}, {and} \bibinfo{person}{Michael~J Black}.} \bibinfo{year}{2024}\natexlab{}.
\newblock \showarticletitle{Wham: Reconstructing world-grounded humans with accurate 3d motion}. In \bibinfo{booktitle}{\emph{Proceedings of the IEEE/CVF Conference on Computer Vision and Pattern Recognition}}. \bibinfo{pages}{2070--2080}.
\newblock


\bibitem[Singh et~al\mbox{.}(2021)]%
        {singh2021real}
\bibfield{author}{\bibinfo{person}{Amritanshu~Kumar Singh}, \bibinfo{person}{Vedant~Arvind Kumbhare}, {and} \bibinfo{person}{K Arthi}.} \bibinfo{year}{2021}\natexlab{}.
\newblock \showarticletitle{Real-time human pose detection and recognition using mediapipe}. In \bibinfo{booktitle}{\emph{International conference on soft computing and signal processing}}. Springer, \bibinfo{pages}{145--154}.
\newblock


\bibitem[Siyao et~al\mbox{.}(2022)]%
        {siyao2022bailando}
\bibfield{author}{\bibinfo{person}{Li Siyao}, \bibinfo{person}{Weijiang Yu}, \bibinfo{person}{Tianpei Gu}, \bibinfo{person}{Chunze Lin}, \bibinfo{person}{Quan Wang}, \bibinfo{person}{Chen Qian}, \bibinfo{person}{Chen~Change Loy}, {and} \bibinfo{person}{Ziwei Liu}.} \bibinfo{year}{2022}\natexlab{}.
\newblock \showarticletitle{Bailando: 3d dance generation by actor-critic gpt with choreographic memory}. In \bibinfo{booktitle}{\emph{Proceedings of the IEEE/CVF Conference on Computer Vision and Pattern Recognition}}. \bibinfo{pages}{11050--11059}.
\newblock


\bibitem[Tarekegn et~al\mbox{.}(2023)]%
        {tarekegn2023enhancing}
\bibfield{author}{\bibinfo{person}{Adane~Nega Tarekegn}, \bibinfo{person}{Mohib Ullah}, \bibinfo{person}{Faouzi~Alaya Cheikh}, {and} \bibinfo{person}{Muhammad Sajjad}.} \bibinfo{year}{2023}\natexlab{}.
\newblock \showarticletitle{Enhancing human activity recognition through sensor fusion and hybrid deep learning model}. In \bibinfo{booktitle}{\emph{2023 IEEE International Conference on Acoustics, Speech, and Signal Processing Workshops (ICASSPW)}}. IEEE, \bibinfo{pages}{1--5}.
\newblock


\bibitem[Tsai et~al\mbox{.}(2023)]%
        {tsai2023automated}
\bibfield{author}{\bibinfo{person}{Ming-Chih Tsai}, \bibinfo{person}{Edward T-H Chu}, {and} \bibinfo{person}{Chia-Rong Lee}.} \bibinfo{year}{2023}\natexlab{}.
\newblock \showarticletitle{An automated sitting posture recognition system utilizing pressure sensors}.
\newblock \bibinfo{journal}{\emph{Sensors}} \bibinfo{volume}{23}, \bibinfo{number}{13} (\bibinfo{year}{2023}), \bibinfo{pages}{5894}.
\newblock


\bibitem[Vermander et~al\mbox{.}(2024)]%
        {vermander2024intelligent}
\bibfield{author}{\bibinfo{person}{Patrick Vermander}, \bibinfo{person}{Aitziber Mancisidor}, \bibinfo{person}{Itziar Cabanes}, {and} \bibinfo{person}{Nerea Perez}.} \bibinfo{year}{2024}\natexlab{}.
\newblock \showarticletitle{Intelligent systems for sitting posture monitoring and anomaly detection: an overview}.
\newblock \bibinfo{journal}{\emph{Journal of neuroengineering and rehabilitation}} \bibinfo{volume}{21}, \bibinfo{number}{1} (\bibinfo{year}{2024}), \bibinfo{pages}{28}.
\newblock


\bibitem[Wallmann-Sperlich et~al\mbox{.}(2013)]%
        {wallmann2013sitting}
\bibfield{author}{\bibinfo{person}{Birgit Wallmann-Sperlich}, \bibinfo{person}{Jens Bucksch}, \bibinfo{person}{Sylvia Hansen}, \bibinfo{person}{Peter Schantz}, {and} \bibinfo{person}{Ingo Froboese}.} \bibinfo{year}{2013}\natexlab{}.
\newblock \showarticletitle{Sitting time in Germany: an analysis of socio-demographic and environmental correlates}.
\newblock \bibinfo{journal}{\emph{BMC public health}}  \bibinfo{volume}{13} (\bibinfo{year}{2013}), \bibinfo{pages}{1--10}.
\newblock


\bibitem[Wang et~al\mbox{.}(2024)]%
        {wang2024double}
\bibfield{author}{\bibinfo{person}{Heng Wang}, \bibinfo{person}{Jieqing Zheng}, \bibinfo{person}{Qin Nie}, \bibinfo{person}{Wenbo Li}, \bibinfo{person}{Zhuo Wang}, \bibinfo{person}{Kun Xiao}, \bibinfo{person}{Xuehao Hu}, \bibinfo{person}{Santosh Kumar}, {and} \bibinfo{person}{Rui Min}.} \bibinfo{year}{2024}\natexlab{}.
\newblock \showarticletitle{Double-Fishtail-Shaped FBG wearable device for sitting posture recognition and real-time respiratory monitoring}.
\newblock \bibinfo{journal}{\emph{IEEE Sensors Journal}} (\bibinfo{year}{2024}).
\newblock


\bibitem[Wu et~al\mbox{.}(2024a)]%
        {wu2024soleposer}
\bibfield{author}{\bibinfo{person}{Erwin Wu}, \bibinfo{person}{Rawal Khirodkar}, \bibinfo{person}{Hideki Koike}, {and} \bibinfo{person}{Kris Kitani}.} \bibinfo{year}{2024}\natexlab{a}.
\newblock \showarticletitle{SolePoser: Full Body Pose Estimation using a Single Pair of Insole Sensor}. In \bibinfo{booktitle}{\emph{Proceedings of the 37th Annual ACM Symposium on User Interface Software and Technology}}. \bibinfo{pages}{1--9}.
\newblock


\bibitem[Wu et~al\mbox{.}(2024b)]%
        {wu2024seeing}
\bibfield{author}{\bibinfo{person}{Ziyu Wu}, \bibinfo{person}{Fangting Xie}, \bibinfo{person}{Yiran Fang}, \bibinfo{person}{Zhen Liang}, \bibinfo{person}{Quan Wan}, \bibinfo{person}{Yufan Xiong}, {and} \bibinfo{person}{Xiaohui Cai}.} \bibinfo{year}{2024}\natexlab{b}.
\newblock \showarticletitle{Seeing through the tactile: 3d human shape estimation from temporal in-bed pressure images}.
\newblock \bibinfo{journal}{\emph{Proceedings of the ACM on Interactive, Mobile, Wearable and Ubiquitous Technologies}} \bibinfo{volume}{8}, \bibinfo{number}{2} (\bibinfo{year}{2024}), \bibinfo{pages}{1--39}.
\newblock


\bibitem[Xu et~al\mbox{.}(2024)]%
        {xu2024mobileposer}
\bibfield{author}{\bibinfo{person}{Vasco Xu}, \bibinfo{person}{Chenfeng Gao}, \bibinfo{person}{Henry Hoffmann}, {and} \bibinfo{person}{Karan Ahuja}.} \bibinfo{year}{2024}\natexlab{}.
\newblock \showarticletitle{Mobileposer: Real-time full-body pose estimation and 3d human translation from imus in mobile consumer devices}. In \bibinfo{booktitle}{\emph{Proceedings of the 37th Annual ACM Symposium on User Interface Software and Technology}}. \bibinfo{pages}{1--11}.
\newblock


\bibitem[Yan and K{\"a}m{\"a}r{\"a}inen(2021)]%
        {yan2021learning}
\bibfield{author}{\bibinfo{person}{Song Yan} {and} \bibinfo{person}{Joni-Kristian K{\"a}m{\"a}r{\"a}inen}.} \bibinfo{year}{2021}\natexlab{}.
\newblock \showarticletitle{Learning anthropometry from rendered humans}.
\newblock \bibinfo{journal}{\emph{arXiv preprint arXiv:2101.02515}} (\bibinfo{year}{2021}).
\newblock


\bibitem[Yu et~al\mbox{.}(2024)]%
        {yu2024seampose}
\bibfield{author}{\bibinfo{person}{Tianhong~Catherine Yu}, \bibinfo{person}{Manru~Mary Zhang}, \bibinfo{person}{Peter He}, \bibinfo{person}{Chi-Jung Lee}, \bibinfo{person}{Cassidy Cheesman}, \bibinfo{person}{Saif Mahmud}, \bibinfo{person}{Ruidong Zhang}, \bibinfo{person}{Fran{\c{c}}ois Guimbreti{\`e}re}, {and} \bibinfo{person}{Cheng Zhang}.} \bibinfo{year}{2024}\natexlab{}.
\newblock \showarticletitle{SeamPose: Repurposing Seams as Capacitive Sensors in a Shirt for Upper-Body Pose Tracking}. In \bibinfo{booktitle}{\emph{Proceedings of the 37th Annual ACM Symposium on User Interface Software and Technology}}. \bibinfo{pages}{1--13}.
\newblock


\bibitem[Zhang et~al\mbox{.}(2024)]%
        {zhang2024dynamic}
\bibfield{author}{\bibinfo{person}{Yu Zhang}, \bibinfo{person}{Songpengcheng Xia}, \bibinfo{person}{Lei Chu}, \bibinfo{person}{Jiarui Yang}, \bibinfo{person}{Qi Wu}, {and} \bibinfo{person}{Ling Pei}.} \bibinfo{year}{2024}\natexlab{}.
\newblock \showarticletitle{Dynamic inertial poser (dynaip): Part-based motion dynamics learning for enhanced human pose estimation with sparse inertial sensors}. In \bibinfo{booktitle}{\emph{Proceedings of the IEEE/CVF Conference on Computer Vision and Pattern Recognition}}. \bibinfo{pages}{1889--1899}.
\newblock


\bibitem[Zhao et~al\mbox{.}(2024)]%
        {zhao20243d}
\bibfield{author}{\bibinfo{person}{Mingjie Zhao}, \bibinfo{person}{Fangting Xie}, \bibinfo{person}{Ziyu Wu}, \bibinfo{person}{Zhen Liang}, {and} \bibinfo{person}{Xiaohui Cai}.} \bibinfo{year}{2024}\natexlab{}.
\newblock \showarticletitle{3D Human Pose Estimation Using Pressure Images on a Smart Chair}. In \bibinfo{booktitle}{\emph{Proceedings of the 2024 2nd Asia Conference on Computer Vision, Image Processing and Pattern Recognition}}. \bibinfo{pages}{1--7}.
\newblock


\bibitem[Zhong et~al\mbox{.}(2024)]%
        {zhong2024accurate}
\bibfield{author}{\bibinfo{person}{Weibing Zhong}, \bibinfo{person}{Hui Xu}, \bibinfo{person}{Yiming Ke}, \bibinfo{person}{Xiaojuan Ming}, \bibinfo{person}{Haiqing Jiang}, \bibinfo{person}{Mufang Li}, {and} \bibinfo{person}{Dong Wang}.} \bibinfo{year}{2024}\natexlab{}.
\newblock \showarticletitle{Accurate and Efficient Sitting Posture Recognition and Human-Machine Interaction Device Based on Fabric Pressure Sensor Array and Neural Network}.
\newblock \bibinfo{journal}{\emph{Advanced Materials Technologies}} \bibinfo{volume}{9}, \bibinfo{number}{3} (\bibinfo{year}{2024}), \bibinfo{pages}{2301579}.
\newblock


\bibitem[Zhou et~al\mbox{.}(2023)]%
        {zhou2023mocapose}
\bibfield{author}{\bibinfo{person}{Bo Zhou}, \bibinfo{person}{Daniel Geissler}, \bibinfo{person}{Marc Faulhaber}, \bibinfo{person}{Clara~Elisabeth Gleiss}, \bibinfo{person}{Esther~Friederike Zahn}, \bibinfo{person}{Lala Shakti~Swarup Ray}, \bibinfo{person}{David Gamarra}, \bibinfo{person}{Vitor~Fortes Rey}, \bibinfo{person}{Sungho Suh}, \bibinfo{person}{Sizhen Bian}, {et~al\mbox{.}}} \bibinfo{year}{2023}\natexlab{}.
\newblock \showarticletitle{Mocapose: Motion capturing with textile-integrated capacitive sensors in loose-fitting smart garments}.
\newblock \bibinfo{journal}{\emph{Proceedings of the ACM on Interactive, Mobile, Wearable and Ubiquitous Technologies}} \bibinfo{volume}{7}, \bibinfo{number}{1} (\bibinfo{year}{2023}), \bibinfo{pages}{1--40}.
\newblock


\bibitem[Zhu et~al\mbox{.}(2023)]%
        {zhu2023motionbert}
\bibfield{author}{\bibinfo{person}{Wentao Zhu}, \bibinfo{person}{Xiaoxuan Ma}, \bibinfo{person}{Zhaoyang Liu}, \bibinfo{person}{Libin Liu}, \bibinfo{person}{Wayne Wu}, {and} \bibinfo{person}{Yizhou Wang}.} \bibinfo{year}{2023}\natexlab{}.
\newblock \showarticletitle{Motionbert: A unified perspective on learning human motion representations}. In \bibinfo{booktitle}{\emph{Proceedings of the IEEE/CVF International Conference on Computer Vision}}. \bibinfo{pages}{15085--15099}.
\newblock


\bibitem[Zhuang et~al\mbox{.}(2023)]%
        {zhuang2023gtn}
\bibfield{author}{\bibinfo{person}{Haolin Zhuang}, \bibinfo{person}{Shun Lei}, \bibinfo{person}{Long Xiao}, \bibinfo{person}{Weiqin Li}, \bibinfo{person}{Liyang Chen}, \bibinfo{person}{Sicheng Yang}, \bibinfo{person}{Zhiyong Wu}, \bibinfo{person}{Shiyin Kang}, {and} \bibinfo{person}{Helen Meng}.} \bibinfo{year}{2023}\natexlab{}.
\newblock \showarticletitle{Gtn-bailando: Genre consistent long-term 3d dance generation based on pre-trained genre token network}. In \bibinfo{booktitle}{\emph{ICASSP 2023-2023 IEEE International Conference on Acoustics, Speech and Signal Processing (ICASSP)}}. IEEE, \bibinfo{pages}{1--5}.
\newblock


\end{thebibliography}


\end{document}